\documentclass[a4paper,aps,prd,10pt,preprintnumbers,showpacs,twocolumn,superscriptaddress,nofootinbib,amsmath,amssymb]{revtex4-1}
\usepackage{times}

\newcommand{\cf}{{cf.}~}
\newcommand{\ie}{{i.e.,}~}
\newcommand{\eg}{{e.g.,}~}

\def\vary{{\cal Y}}

\begin{document}
\title{General parametrization of axisymmetric black holes in metric
  theories of gravity}

\author{Roman Konoplya}
\affiliation{Institute for Theoretical Physics, Goethe University,
  Max-von-Laue-Str. 1, 60438 Frankfurt, Germany}

\author{Luciano Rezzolla}
\affiliation{Institute for Theoretical Physics, Goethe University,
  Max-von-Laue-Str. 1, 60438 Frankfurt, Germany}
\affiliation{Frankfurt Institute for Advanced Studies, Goethe-Universit\"at,
  Ruth-Moufang-Str. 1, 60438 Frankfurt, Germany}

\author{Alexander Zhidenko}
\affiliation{Institute for Theoretical Physics, Goethe University,
  Max-von-Laue-Str. 1, 60438 Frankfurt, Germany}
\affiliation{Centro de Matem\'atica, Computa\c{c}\~ao e Cogni\c{c}\~ao,
  Universidade Federal do ABC (UFABC), Rua Aboli\c{c}\~ao, CEP:
  09210-180, Santo Andr\'e, SP, Brazil}

\begin{abstract}
  Following previous work of ours in spherical symmetry, we here propose a new parametric framework to describe the spacetime of axisymmetric black holes in generic metric theories of gravity. In this case, the metric components are functions of both the radial and the polar angular coordinates, forcing a double expansion to obtain a generic axisymmetric metric expression. In particular, we use a continued-fraction expansion in terms of a compactified radial coordinate to express the radial dependence, while we exploit a Taylor expansion in terms of the cosine of the polar angle for the polar dependence. These choices lead to a superior convergence in the radial direction and to an exact limit on the equatorial plane.  As a validation of our approach, we build parametrized representations of Kerr, rotating dilaton, and Einstein-dilaton-Gauss-Bonnet black holes. The match is already very good at lowest order in the expansion and improves as new orders are added. We expect a similar behavior for any stationary and axisymmetric black-hole metric.
\end{abstract}

\maketitle
\section{Introduction}

The existence of an event horizon would be indisputable if it was
obtained by the direct observation of gravitational waves from a
perturbed black hole. This is because the gravitational response of a
perturbed black hole would be intrinsically different from that of another
putative compact object who would have essentially the same properties in
terms of electromagnetic emission (see Refs. \cite{Mazur:05,Chirenti:06}
for the case of gravastars).

Properties of the event horizon could probably be studied through
analysis of an electromagnetic spectrum of the accreting matter
\cite{Abramowicz:02}. The radio compact source Sgr~A$^*$, which is
assumed to be a supermassive black hole at the center of our Galaxy, is
the best option for such investigation of the event horizon.  Recent
radio observations of Sgr~A$^*$ achieved scales comparable to what should
be the size of the event horizon \cite{Doeleman}. In the near future,
very long baseline interferometric radio observations are expected to
image the so-called black-hole ``shadow''~\cite{Falcke:00} -- the photon
ring marking the surface where photons will have their smallest stable
orbit \cite{Johannsen:2012vz,Abdujabbarov:15}. In addition to providing
the evidence for the existence of black holes, these observations could
also help in testing the no-hair theorem in general relativity
\cite{Bambi:2008jg,JohannsenPsaltis:2010,Loeb:2013lfa,Psaltis:2015uza} as
well as testing of general relativity itself against a number of
alternative theories of gravity.

Because of the large number of alternative theories of gravity and some
possibility that the ``true'' theory is yet unknown, it is reasonable to
develop a model-independent framework which parametrizes the most generic
black-hole geometry through a finite number of adjustable
quantities. These quantities must be chosen in such a way that they can
be used to measure deviations from the general-relativistic black-hole
geometry (Kerr metric) and could be estimated from the observational data
\cite{Vigeland:2011ji}. This approach is similar in spirit to the
parametrized post-Newtonian approach (PPN) which describes the spacetime
far from the source of strong gravity \cite{Will:2005va}.

One of the first such parametrizations for black holes was proposed by
Johannsen and Psaltis~\cite{Johannsen:2011dh}, who expressed deviations
from general relativity in terms of a Taylor expansion in powers of
$M/r$, where $M$ and $r$ are the mass of the black hole and a generic
radial coordinate. While some of the first coefficients of the expansion
can be easily constrained in terms of PPN-like parameters, an infinite
number remains to be determined from observations near the event horizon
\cite{Johannsen:2011dh}. As pointed out by Ref. \cite{Cardoso:2014rha},
this approach faces a number of difficulties:

\begin{itemize}
\item[\emph{(i)}] A generic metric would be described by an infinite
  number of roughly equally important parameters, making it difficult to
  isolate the dominant ones.

\item[\emph{(ii)}] The parametrization can be employed to study small
  deviations from general relativity, but fails for essentially
  non-Einsteinian theories of gravity, such as, for example,
  Einstein-dilaton-Gauss-Bonnet (EDGB) gravity with large coupling
  constant \cite{Cardoso:2014rha}.

\item[\emph{(iii)}] The adoption of the Janis-Newman transformations
  \cite{Janis:65} in the Johannsen-Psaltis parametrization
  \cite{Johannsen:2011dh} does not allow one to reproduce alternatives to
  the Kerr spacetime, for examples in dilaton or Chern-Simons modified
  gravity.

\end{itemize}
As a result, despite the best intentions and the wide adoption of this
metric, the Johannsen-Psaltis approach basically does not seem to be a
robust and generic parametrization for rotating black holes.

In a previous paper of ours \cite{RezzollaZhidenko}, the solution to the
above issues was proposed for arbitrary spherically symmetric black holes
in metric theories of gravity.  This was achieved by expressing the
deviations from general relativity in terms of a continued-fraction
expansion via a compactified radial coordinate defined between the event
horizon and spatial infinity. The superior convergence properties of this
expansion effectively reduced to a few the number of coefficients
necessary to approximate such spherically symmetric metric to the
precision that can be in principle probed with near-future observations.

In this paper we extend our approach to arbitrary axially symmetric
spacetimes, describing rotating black holes in metric theories of
gravity. In particular, using an asymptotic spherical coordinate system
and starting from the metric functions as expressed on the equatorial
plane, we perform an expansion in terms of powers of $\cos \theta$, where
$\theta$ is the polar coordinate. Under appropriate choice of
coordinates, leading to the Boyer-Lindquist ones in the case of zero
deviations from the Kerr geometry, the expansion in $\cos \theta$ in the
polar direction provides excellent convergence for known black-hole
metrics, such as rotating Einstein-dilaton-Gauss-Bonnet and
Johannsen-Psaltis ones.

For a number of other cases, such as Kerr, Kerr-Newman, Sen and others,
the expansions in terms of powers of $\cos \theta$ converge to the
corresponding exact solutions already at the second order. At the same
time, when the metric components are expressed as rational functions of
the radial coordinate $r$, the expansion in the radial direction follows
the same behavior discussed in \cite{RezzollaZhidenko} and thus
converges rapidly with a finite and small number of terms.  As a result,
with the approach introduced here, a number of black-hole metrics,
including the Kerr spacetime, are reproduced \emph{``exactly''} and in
the whole space, \ie from the event horizon out to spatial infinity.

The paper is organized as follows: In Sec. \ref{sec:abhs} we describe a
general ansatz for axisymmetric black holes, which is further constrained
by a specific coordinate choice developed in
Sec. \ref{sec:ccaeitpd}. Section \ref{sec:parametrization} is devoted to
construction of the generic parametrization, while in
Sec. \ref{sec:circulargeo} we present the general procedure for
calculation of multipole moments through the properties of the circular
geodesic motion of particles around black holes. Section \ref{sec:dilbh}
is devoted to construction of the parametrization for a dilaton rotating
black hole and we illustrate there how the parametrization has superior
convergence by computing the binding energy for exact and parametrized
(at different orders) dilaton black-hole spacetimes. Section
\ref{sec:EDGBbh} presents instead the parametrization for an
Einstein-dilaton-Gauss-Bonnet (EDGB) rotating black hole.  Finally,
Sec. \ref{sec:conclusions} summarizes our results and presents our
conclusions. To facilitate the use of our parametrized metrics, Appendix
\ref{sec:appendix_a} provides a collection of the explicit expressions of
the parametrized metrics for a Kerr, dilaton, and EDGB black hole.

\section{Axisymmetric black holes}
\label{sec:abhs}

The general form of an axisymmetric line element allows the coordinates
$t$ and $\phi$ to be along the direction selected by two Killing vectors
that are timelike and spacelike, respectively. It is convenient to choose
the other two spacelike coordinates, $\rho$ and $\vartheta$ to be
mutually orthogonal and orthogonal to the coordinates $t$ and $\phi$,
such that ($\rho$, $\theta$, $\phi$) are spherical coordinates at spatial
infinity. In this way the general form of the metric tensor for axially
symmetric spacetimes can be written as
\begin{widetext}
\begin{equation}
\label{axisymmetric}
ds^2 =
-\dfrac{f(\rho,\vartheta)-
\omega^2(\rho,\vartheta)\sin^2\vartheta}{\kappa^2(\rho,\vartheta)}
dt^2
-2\omega(\rho,\vartheta)\rho\sin^2\vartheta
dtd\phi+\kappa^2(\rho,\vartheta)\rho^2\sin^2\vartheta
d\phi^2
+\sigma(\rho,\vartheta)\left(\dfrac{\beta^2(\rho,\vartheta)}
{f(\rho,\vartheta)}d\rho^2+\rho^2d\vartheta^2\right)\,,
\end{equation}
\end{widetext}
where $f(\rho,\vartheta)$, $\beta(\rho,\vartheta)$,
$\sigma(\rho,\vartheta)$, $\kappa(\rho,\vartheta)$, and
$\omega(\rho,\vartheta)$ are some dimensionless functions of the two
coordinates $\rho$ and $\vartheta$.

A generic axisymmetric black-hole spacetime with line element expressed
by (\ref{axisymmetric}) is expected to have a compact axisymmetric event
horizon. Outside the horizon the introduced functions
$f(\rho,\vartheta)$, $\beta(\rho,\vartheta)$, $\sigma(\rho,\vartheta)$,
$\kappa(\rho,\vartheta)$ are finite and positive definite, so as to avoid
any ``metric issue'', such as singularities, closed time-like
trajectories, etc.

Given the metric \eqref{axisymmetric}, it is easy to find that
\begin{equation}
\sqrt{-g} \equiv
\sqrt{-\det(g_{\mu\nu})} =
\beta(\rho,\vartheta)\sigma(\rho,\vartheta)\rho^2\sin\vartheta\,,
\end{equation}
and
\begin{equation}
g^{\rho\rho} = \frac{f(\rho,\vartheta)}{\beta^2(\rho,\vartheta)}\,.
\end{equation}

From the latter expression we conclude that the Killing horizon (\ie the
surface defined by the null Killing vector) is given by
\begin{equation}
f(\rho,\vartheta) = 0\,,
\end{equation}
while the ergoregion is instead defined as
\begin{equation}
\label{ergoregion}
0< f(\rho,\vartheta)<\omega^2(\rho,\vartheta)\sin^2\vartheta\,.
\end{equation}

Since $\omega(\rho,\vartheta)$ is finite at the Killing horizon, the
ergosphere always touches the horizon at the poles $\vartheta = 0,\pi$.

Further we shall consider the rotating dilaton black hole (or Kerr-Sen)
black hole \cite{Sen} or axion-dilaton black hole with zero
Newman-Unti-Tamburino (NUT) charge \cite{dilaton-BH}) as one of the examples.
The above five functions for this black hole are
\begin{subequations}
\label{dilatonmetric}
\begin{eqnarray}
\label{dilatonmetric_1}
f(\rho,\vartheta)& = &\frac{\rho^2-2\mu\rho+a^2}{\rho^2}\,,\\
\label{dilatonmetric_2}
\beta(\rho,\vartheta)& = &1\,,\\
\label{dilatonmetric_3}
\sigma(\rho,\vartheta)& = &\frac{\rho^2+2b\rho+a^2\cos^2\vartheta}{\rho^2}\,,\\
\label{dilatonmetric_4}
\kappa^2(\rho,\vartheta)& =
&\frac{(\rho^2+2b\rho+a^2)^2-a^2\sin^2\vartheta
(\rho^2-2\mu\rho+a^2)}{\rho^2(\rho^2+2b\rho+a^2\cos^2\vartheta)}\,,\nonumber\\\\
\label{dilatonmetric_5}
\omega(\rho,\vartheta)& = &\frac{2(\mu+b)a}{\rho^2+2b\rho+a^2\cos^2\vartheta}\,,
\end{eqnarray}
\end{subequations}
where $a \equiv J/M$ and $b$ are parameters of rotation and dilaton
respectively, and $\mu \equiv M- b$, where $M$ is the black-hole
mass. Clearly, when $b = 0$, the above metric components reduce to those
of the Kerr spacetime in Boyer-Lindquist coordinates $(\rho, \theta,
\phi)$.

\section{Coordinate choices and expansion in the polar direction}
\label{sec:ccaeitpd}

The choice of functions $f(\rho,\vartheta)$, $\beta(\rho,\vartheta)$,
$\sigma(\rho,\vartheta)$, $\kappa(\rho,\vartheta)$, and
$\omega(\rho,\vartheta)$ for a given axisymmetric black hole is not
unique. Instead of $\rho$ and $\vartheta$ we could in fact choose another
couple of coordinates, which are also mutually orthogonal and orthogonal
to the coordinates $t$ and $\phi$. Thus, at this stage the coordinates
are not fully fixed: a single black hole can be represented by a number
of different systems of coordinates within the same initial ansatz
(\ref{axisymmetric}). In the end this would produce a nonunique
parametrization, when the same black hole could be described by a number
of different sets of parameters. To avoid such a degeneracy, we will here
impose further conditions which fix the coordinates fully.

In order to describe how to attain such a unique coordinate choice, let
us first consider transformations from the initial coordinates $(\rho,
\vartheta)$ to the new ones $(r, \theta)$, such that the new line element
is expressed in terms of five metric functions $N(r,\theta)$,
$W(r,\theta)$, $K(r,\theta)$, $B(r,\theta)$, $\Sigma(r,\theta)$, and has
the following form
\begin{widetext}
\begin{equation}
\label{fixedmetric}
ds^2 =
-\dfrac{N^2(r,\theta)-W^2(r,\theta)\sin^2\theta}{K^2(r,\theta)}dt^2
-2W(r,\theta)r\sin^2\theta dtd\phi+K^2(r,\theta)r^2\sin^2\theta d\phi^2
+\Sigma(r,\theta)\left(\dfrac{B^2(r,\theta)}{N^2(r,\theta)}dr^2 +
r^2d\theta^2\right)\,.
\end{equation}
\end{widetext}

Comparing the inverse metric components of (\ref{axisymmetric}) and
(\ref{fixedmetric}) we find the following relations
\begin{subequations}
\label{coordtransform}
\begin{eqnarray}
N^2(r,\theta)r^2\sin^2\theta \!&=&\!
  f(\rho,\vartheta)\rho^2\sin^2\vartheta\,,\label{eqN}\\
W(r,\theta)r\sin^2\theta \!&=&\!
  \omega(\rho,\vartheta)\rho\sin^2\vartheta\,,\label{eqW}\\
K^2(r,\theta)r^2\sin^2\theta \!&=&\!
  \kappa^2(\rho,\vartheta)\rho^2\sin^2\vartheta\,,\label{eqK}\\
\dfrac{N^2(r,\theta)}{\Sigma(r,\theta)B^2(r,\theta)} \!&=&\!
  \dfrac{1}{\sigma(\rho,\vartheta)}\!\!\left(
  \dfrac{f(\rho,\vartheta)}{\beta^2(\rho,\vartheta)}\dfrac{\partial
    r}{\partial \rho}\dfrac{\partial r}{\partial
    \rho}\!+\!\dfrac{1}{\rho^2}\dfrac{\partial r}{\partial
    \vartheta}\dfrac{\partial r}{\partial
    \vartheta}\right)\,,\nonumber\\\label{eqB}\\
\dfrac{1}{\Sigma(r,\theta)r^2} \!&=&\! \dfrac{1}{\sigma(\rho,\vartheta)}\!\!\left(
\dfrac{f(\rho,\vartheta)}{\beta^2(\rho,\vartheta)}\dfrac{\partial
  \theta}{\partial \rho}\dfrac{\partial \theta}{\partial
  \rho}\!+\!\dfrac{1}{\rho^2}\dfrac{\partial \theta}{\partial
  \vartheta}\dfrac{\partial \theta}{\partial
  \vartheta}\right)\,,\label{eqR}\nonumber\\\\
0 \!&=&\! \dfrac{f(\rho,\vartheta)}{\beta^2(\rho,\vartheta)}\dfrac{\partial
    r}{\partial \rho}\dfrac{\partial \theta}{\partial
    \rho}+\dfrac{1}{\rho^2}\dfrac{\partial r}{\partial
    \vartheta}\dfrac{\partial \theta}{\partial \vartheta}\,.\label{eqT}
\end{eqnarray}
\end{subequations}

The last two equations define the relation between the coordinates $(r,
\theta)$ and $(\rho, \vartheta)$. Once one finds $r$ and $\theta$ as
functions of $\rho$ and $\vartheta$ through Eqs. (\ref{eqR}),
(\ref{eqT}), it is possible to find the functions $N(r,\theta)$,
$W(r,\theta)$, $K(r,\theta)$, and $B(r,\theta)$ from Eqs. (\ref{eqN}),
(\ref{eqW}), (\ref{eqK}), and (\ref{eqB}), respectively.

Next, we assume that the functions $f(\rho,\vartheta)$,
$\beta(\rho,\vartheta)$, $\sigma(\rho,\vartheta)$,
$\kappa(\rho,\vartheta)$, and $\omega(\rho,\vartheta)$ are known as
series expansion in terms of small parameter $\vary$ measuring the
distance from the equatorial plane, \ie defined as
$\vary\equiv\cos\vartheta$. Hence, we have

\begin{subequations}
\label{expansionY}
\begin{eqnarray}
f &=&
f_0(\rho)+f_1(\rho)\,\vary+f_2(\rho)\,\vary^2+\mathcal{O}(\vary^3)\,\\
\beta &=&
\beta_0(\rho)+\beta_1(\rho)\,\vary+\beta_2(\rho)\,\vary^2+\mathcal{O}(\vary^3)\,\\
\sigma &=&
\sigma_0(\rho)+\sigma_1(\rho)\,\vary+\sigma_2(\rho)\,\vary^2+\mathcal{O}(\vary^3)\,\\
\kappa &=&
\kappa_0(\rho)+\kappa_1(\rho)\,\vary+\kappa_2(\rho)\,\vary^2+\mathcal{O}(\vary^3)\,\\
\omega &=&
\omega_0(\rho)+\omega_1(\rho)\,\vary+\omega_2(\rho)\,\vary^2+\mathcal{O}(\vary^3)\,
\end{eqnarray}
\end{subequations}
where the coefficients in the expansion, \ie $f_i, \beta_i, \sigma_i,
\ldots$ are functions of the radial coordinate only.

Introducing now the small parameter $y\equiv\cos\theta$, we can find the
corresponding new transformation $(\vary, \rho) \rightarrow (y, r)$ as a
series expansion in terms of $\vary$, namely
\begin{eqnarray}
\label{transform-coord}
y & = &\vary(1+\alpha_0(\rho))\left[1+\alpha_1(\rho)\vary +
\alpha_2(\rho)\vary^2+\mathcal{O}(\vary^2)\right]\,,\nonumber\\\\
r^2& = &\rho^2(1+\zeta_0(\rho))\left[1+\zeta_1(\rho)\vary +
\zeta_2(\rho)\vary^2+\mathcal{O}(\vary^2)\right]\,.\nonumber\\
\end{eqnarray}
A few remarks should be made here. First, if all the coefficients
$\alpha_i(\rho) = 0$, then $\theta =
\vartheta$, that is, the two polar coordinates are identical. Second, if
all coefficients $\zeta_i(\rho) = 0$, then also the two radial
coordinates are identical, \ie $r = \rho$. Finally and most important,
we have here decided to perform a generic expansion in powers of the
small parameter $\cos\theta$. This choice allows us to consider black
holes that are not reflection symmetric across the equatorial plane. This
is admittedly a bizarre possibility, but one we want to preserve for
generality. In practice, any representation of an astrophysically
reasonable black hole would require also the reflection symmetry across
the equatorial plane, thus limiting the expansion to \emph{even} powers
of $\cos\theta$. Indeed this is what we will do when considering the
parametrization of Kerr, dilaton, and EDGB black holes in the following
sections.

Without loss of generality we can assume that there are observers at
spatial infinity that are able to measure the angular momentum $J \equiv
A M$ and mass $M$ of the black hole. With this assumption, we are then
able to define the coordinates $r$ and $\theta$ unambiguous.

More specifically, we first need to fix one of the five metric functions
$N(r,\theta)$, $W(r,\theta)$, $K(r,\theta)$, $B(r,\theta)$, $\Sigma(r,
\theta)$. Out of the five possible choices, we prefer to fix the function
$\Sigma (r, \theta)$ as
\begin{equation}
\label{Sigmacondition}
\Sigma(r, \theta) = 1+\frac{A^2y^2}{r^2} = 1+\frac{A^2\cos^2\theta}{r^2}\,,
\end{equation}
since this allows us to reproduce the Boyer-Lindquist coordinates for the
case of a Kerr black hole.  This choice is also compatible with
asymptotic behavior of the asymptotically flat and axially symmetric
metric of a rotating body in the Boyer-Lindquist coordinates
\begin{eqnarray}
d s^2 &\approx& - \left(1- \frac{2 M}{r}\right) d t^2- \frac{4 M A \sin^2
  \theta}{r} d t\, d \phi + dr^2 \nonumber\\&&+ r^2(d\theta^2+\sin^2\theta
d\phi^2)\,.
\label{BL}
\end{eqnarray}

In an astrophysically realistic context, at large distance the
gravitational field of a rotating object of any kind should be
essentially Newtonian, thus, hiding the details related to the event
horizon of an isolated black hole and regime of strong gravity. The
addition of extra fields to the black-hole spacetime, \eg scalar or
electromagnetic, would certainly change the asymptotic behavior. Yet,
backreaction of such fields onto the background geometry is expected to
be negligibly small for astrophysical black holes, thereby, allowing us
to consider an isolated black hole as in vacuum.

A more careful analysis reveals that the condition (\ref{Sigmacondition})
is still insufficient to fix completely the freedom of the coordinate
choice. The reason for this is that using Eqs. (\ref{eqR}) and
(\ref{eqT}) it is possible to obtain different series for $y$ and $r$ for
each different choice of $\alpha_0(\rho)$. Thus, the function
$\alpha_0(\rho)$ must be fixed in order to achieve the uniqueness of the
coordinate transformations (\ref{transform-coord}).  The natural way to
fix $\alpha_0(\rho)$ is to choose an additional condition for the new
line element in the equatorial plane in such a way that for the zero
rotation we reproduce spherical coordinates. This can be done in various
ways, but we here choose to impose a condition on the function $K$ on the
equatorial plane. More specifically, we first observe that, multiplying
(\ref{eqW}) by any constant $C$ and adding (\ref{eqK}), we obtain
\begin{equation}
\left(K^2+\dfrac{C}{r}W\right)r^2\sin^2\theta =
\left(\kappa^2+\dfrac{C}{r}\omega\right)\rho^2\sin^2\vartheta\,,
\end{equation}
which, in turn, allows us to impose that
\begin{equation}\label{Kcondition}
\left(K^2-\dfrac{A}{r}W\right) = 1+\frac{A^2}{r^2}+{\cal O}(y)\,.
\end{equation}
It is not difficult to verify that the Boyer-Lindquist coordinates fulfill
the above condition. This gives us the unambiguous coordinate choice for
$r$ and $\theta$.

In summary, in order to transform unambiguously any given coordinates
$(\rho,\vartheta)$ to the new coordinates $(r,\theta)$ we need to
\begin{enumerate}
\item define the rotation parameter $A \equiv J/M$, where $J$ is the
  total angular momentum and $M$ is the asymptotic mass of the
  spacetime\footnote{The constant $M$ could be associated with the Arnowitt-Deser-Misner (ADM)
    mass at spatial infinity if such a quantity can be properly
    defined. However, because we are not limiting ourselves to
    asymptotically flat spacetimes where such a quantity is defined, we
    here consider the more general case in which astronomical
    observations at large distances from the event horizon but not at
    spatial infinity can be exploited to measure the constant $M$.};

\item choose $r$ and $\theta$ to be mutually orthogonal and orthogonal
  to the coordinates $t$ and $\phi$;

\item impose that the metric functions satisfy the conditions
  (\ref{Sigmacondition})~and~(\ref{Kcondition}), namely
\begin{eqnarray}
\label{Sigma_of_y}
\Sigma(r,\theta) &=& 1+\frac{A^2}{r^2}\cos^2\theta\,,\\
\label{K_of_y}
K^2\left(r,\frac\pi2\right)-\dfrac{A}{r}W\left(r,\frac\pi2\right) &=&
1+\frac{A^2}{r^2}\,.
\end{eqnarray}
\end{enumerate}
The latter condition in fact allows us to fix the function
$\alpha_0(\rho)$ as
\begin{equation}\label{alpha0}
1+\alpha_0(\rho) =
\sqrt{\dfrac{\sigma_0(\rho)}{\kappa_0^2(\rho)-
A\omega_0(\rho)/\rho-A^2/\rho^2}}\,.
\end{equation}
so that, from Eqs. (\ref{eqR}) and (\ref{eqT}), we finally find
\begin{subequations}
\label{yseries}
\begin{eqnarray}
y &=&\vary\left(1+\alpha_0(\rho)\right)
\left[1+\frac{\sigma_1(\rho)}{4\sigma_0(\rho)}\vary+{\cal
  O}(\vary^2)\right]\,,\\
r^2 &=&
\frac{\rho^2\sigma_0(\rho)}{(1+\alpha_0(\rho))^2}+{\cal O}(\vary^2)\,.
\end{eqnarray}
\end{subequations}

In analogy with what was done for the expressions (\ref{expansionY}), we can
now invert the series (\ref{yseries}) to find the
functions $\vary(r,y)$ and $\rho(r,y)$ and thus express the metric
functions $N^2$, $W$, $K^2$, and $B^2$ in the line element
(\ref{fixedmetric}) as a series in terms of the new variable $y$.

\section{Parametrization for axisymmetric black holes}
\label{sec:parametrization}

In order to obtain the proper asymptotic behavior of the newly
introduced metric function we need to ensure they satisfy the following
behavior at large distances, namely that for $r \gg 1$
\begin{subequations}
\label{asymptot}
  \begin{eqnarray}
    N^2(r,\theta)& = &1-\dfrac{2M}{r}+{\cal O}\left(\dfrac{1}{r^2}\right)\,,\\
    B(r,\theta)& = &1+{\cal O}\left(\dfrac{1}{r}\right)\,,\\
    W(r,\theta)& = &{\cal O}\left(\dfrac{1}{r^2}\right)\,,\\
    K^2(r,\theta)& = &1+{\cal O}\left(\dfrac{1}{r^2}\right)\,,
  \end{eqnarray}
\end{subequations}
where $M$ is a constant to be read at a large distance.

These conditions on the metric functions obviously imply that the metric
is asymptotically flat and spherically symmetric at large distance from
the black hole. In principle, these conditions could be violated in a
cosmology admitting violation of isotropy, such as the Einstein-Aether
theory. Yet, it is evident that local physical processes around black
holes cannot be influenced by such cosmological factors while the
coupling constant of new interactions (be it vector Aether or any other
field) is negligibly small for observations of localized processes.

Now that we have performed the coordinate transformations (\ref{yseries})
and all the functions are obtained as series expansion in terms of the
``angular'' variable $y = \cos\theta$, it is necessary to introduce the
parametrization for the coefficients of the series which are functions of
the radial coordinate only. Also in this case there are several different
ways in which this can be accomplished. Here, however, we will follow the
powerful approach already employed in \cite{RezzollaZhidenko} for the
parametrization of a generic black-hole metric in spherical
symmetry. More specifically, we first introduce the compact coordinate
radial
\begin{equation}
x = 1-\frac{r_0}{r}\,,
\end{equation}
where $r_0$ is the black-hole horizon radius in the equatorial plane; \ie
$r_0$ is the largest solution of the equation
\begin{equation}
N^2(r,\pi/2) = 0\,.
\end{equation}
Clearly $x\in [0,1]$, with $x=0$ at the black-hole horizon on the
equatorial plane (\ie $y=0$) and $x=1$ at spatial infinity.

Second, we consider the following expansions in terms of the new compact
coordinate $x$,
\begin{subequations}
\label{yexp}
\begin{eqnarray}
N^2 &=& xA_0(x)+\sum\limits_{i = 1}^{\infty}A_i(x)y^i\,, \\
B &=& 1+\sum\limits_{i = 0}^{\infty}B_i(x)y^i\,,\\
W &=& \sum\limits_{i = 0}^{\infty}\dfrac{W_i(x)y^i}{\Sigma}\,, \\
K^2-\dfrac{AW}{r} &=& 1+\sum\limits_{i = 0}^{\infty}\dfrac{K_i(x)y^i}{\Sigma}\,,
\end{eqnarray}
\end{subequations}
where from \eqref{Sigma_of_y}
\begin{equation}
\Sigma = 1+\dfrac{A^2}{r_0^2}(1-x)^2y^2\,.
\end{equation}
Our coordinate choice (\ref{Kcondition}) then fixes $K_0$ to be
\begin{equation}\label{coordcond}
K_0(x) = (1-x)^2\frac{A^2}{r_0^2}\,.
\end{equation}

Third, in order to satisfy the required asymptotic behavior
(\ref{asymptot}), we define
\begin{subequations}
\label{fdef}
\begin{eqnarray}
\label{bdef}
B_i(x) &=& b_{i0}(1-x)+{\tilde B}_i(x)(1-x)^2\,,\\
\nonumber \\
\label{wdef}
W_i(x) &=& w_{i0}(1-x)^2+{\tilde W}_i(x)(1-x)^3\,,\\
\nonumber \\
\label{kdef}
K_i(x) &=& k_{i0}(1-x)^2+{\tilde K}_i(x)(1-x)^3\,,\\
\nonumber \\
\label{a0def}
A_0(x) &=& 1-\epsilon_0(1-x)+(a_{00}-\epsilon_0+k_{00})(1-x)^2
\nonumber \\
&& \phantom{1} +{\tilde A}_0(x)(1-x)^3\,,\\
\nonumber \\
\label{aidef}
A_{i>0}(x) &=& K_i(x)+\epsilon_{i}(1-x)^2+a_{i0}(1-x)^3+\nonumber\\
&& \phantom{K_i(x)}+{\tilde A}_i(x)(1-x)^4\,.
\end{eqnarray}
\end{subequations}
Note that the coefficients $\epsilon_i$, $a_{i0}$, $b_{i0}$, $w_{i0}$,
$k_{i0}$ for $i = 0,1,2,3\ldots$ are fixed in such a way that the
expansion (\ref{fdef}) matches desired asymptotic behavior near spatial
infinity (\ie $x = 1$). To make it astrophysically meaningful, the latter
should be expressed in the terms of the PPN expansion.

Since the tilted functions ${\tilde A}_i(x)$, ${\tilde B}_i(x)$, ${\tilde
  W}_i(x)$, and ${\tilde K}_i(x)$ describe the black-hole metric near its
horizon, we here express them in close analogy with what was already done in
\cite{RezzollaZhidenko}, and thus parametrize them in terms of Pad\'e
approximants in the form of continued fraction
\begin{subequations}
\label{tiltedfunctions}
\begin{eqnarray}
{\tilde A}_i(x)& = &\dfrac{a_{i1}}{1+\dfrac{a_{i2}x}{1+\dfrac{a_{i3}x}{1+\ldots}}}\,,\\
\nonumber \\
{\tilde B}_i(x)& = &\dfrac{b_{i1}}{1+\dfrac{b_{i2}x}{1+\dfrac{b_{i3}x}{1+\ldots}}}\,,\\
\nonumber \\
{\tilde W}_i(x)& = &\dfrac{w_{i1}}{1+\dfrac{w_{i2}x}{1+\dfrac{w_{i3}x}{1+\ldots}}}\,,\\
\nonumber \\
{\tilde K}_i(x)& = &\dfrac{k_{i1}}{1+\dfrac{k_{i2}x}{1+\dfrac{k_{i3}x}{1+\ldots}}}\,,
\end{eqnarray}
\end{subequations}
where $a_{ij}$, $b_{ij}$, $w_{ij}$, $k_{ij}$ for $i = 0,1,2,3\ldots$, $j
= 1,2,3\ldots$ are fixed via a comparison of the series expansions of the
metric functions near the black-hole horizon (\ie $x = 0$) with their
exact analytic expressions or of numerical data when an exact solution
cannot be obtained analytically.

To recap: Two different sets of coefficients appear in the approach
proposed here for the parametrization of a generic stationary and
axisymmetric black-hole metric. The first set is given by the
coefficients $a_{i0}$, $b_{i0}$, $w_{i0}$, $k_{i0}$, $\epsilon_{i}$ ($i =
1,2,3\ldots$), which are fixed by spacetime behavior in the asymptotic
region, \ie $x \to 1^{-}$. The second set is instead given by the
coefficients $a_{ij}$, $b_{ij}$, $w_{ij}$, $k_{ij}$ ($j = 1,2,3\ldots$),
which are fixed by the geometry of the black hole near its horizon, \ie
$x \to 0^{+}$. Needless to say, such a separation of parameters on ``near
horizon'' and ``asymptotic'' is essential for the comparison of the
theoretical predictions with the observational data obtained, either in
the far region (\eg values of PPN parameters) or near the black hole (\eg
study of accretion flows, black hole's shadows, etc.). It should also be
noted that the contribution of $K_i(x)$ in the definition of $A_{i>0}(x)$
(\ref{aidef}) allows us to define the asymptotic parameters
$\epsilon_{i}$ and $a_{i0}$, by comparing the asymptotic expansions of
the metric component $g_{tt}$. Without this contribution of $K_i(x)$, we
would not be able to separate the parameters $a_{i0}$ and $k_{ij}$ into,
``asymptotic'' and ``near-horizon'' ones, respectively.

It is important to discuss now some essential properties of the continued
fraction used in the Pad\'e approximation in (\ref{tiltedfunctions}),
namely, the number of terms $N$ appearing in the continued fractions.  To
this scope we will consider three different cases.

The first case is the one for which the tilted functions ${\tilde
  A}_i(x)$, ${\tilde B}_i(x)$, ${\tilde W}_i(x)$, and ${\tilde K}_i(x)$
are fractions of two polynomials of $r$, as it happens, for example, for
any of these functions in the case of the Kerr solution. In this case,
then, the corresponding expansion (\ref{tiltedfunctions}) contains a
\emph{finite} and \emph{small} number of terms. In other words, there
exist a number $N>0$, such that $a_{iN} = 0$, or $b_{iN} = 0$, or $w_{iN}
= 0$, or $k_{iN} = 0$, respectively. In this case, all higher-order terms
are obviously not necessary and we refer to this representation as
\emph{``exact''}, in the sense that the corresponding metric can be
reproduced analytically with only a finite number of coefficients.

The second case is for when the metric functions $N(r, \theta)$, $W(r,
\theta)$, $K(r, \theta)$, and $B(r, \theta)$ in (\ref{fixedmetric}) are
not rational functions of $r$. In this case, then, some or all of the
expansions (\ref{tiltedfunctions}) contain an infinite number of
coefficients, \ie $N = \infty$. Finally, the third case is for when even
though $N$ is finite, there is no guarantee that an ``exact''
representation will be achieved with a small number of coefficients. As a
result, in both of these last two cases (\ie of finite but large and
infinite $N$), the metric representation will be only approximate and
limited to the first $n<N$ terms, setting the $n$-th equal to zero. We
will refer to such approximated metrics obtained by truncating at the
$n$-th term as the ``approximation of the $(n-1)$-th order''.

As a corollary to the previous remark we also note that the $n$-th
coefficient cannot always be set equal to zero in a consistent manner. In
some cases, in fact, setting a particular coefficient to zero could imply
the divergence of the truncated continued fractions
(\ref{tiltedfunctions}) for the corresponding tilted function for some
value of $r$ outside the event horizon. To solve this problem within the
approximation of the $(n-1)$-th order, one should set the $(n+1)$-th
coefficient equal to zero, and choose an arbitrary value for the $n$-th
coefficient, so that the denominator remains positive definite for all
values of the radial coordinate outside the event horizon (\ie for
$x>0$). For convenience, in such cases we will take the value of the
$n$-th coefficient to be equal to unity and will refer to these cases as
approximations of $(n-1)$-th order as well.

As a final remark we note that not all of the parameters so far, \ie
$a_{ij}$, $b_{ij}$, $w_{ij}$, and $k_{ij}$, are effectively
independent. This is because one of the infinite number of functions, \ie
$A_i(x)$, $B_i(x)$, $W_i(x)$, and $K_i(x)$, must be fixed by a coordinate
choice. We have here used the condition (\ref{coordcond}), so that,
taking into account the continued fractions (\ref{tiltedfunctions}),
yields
\begin{equation}
\label{coor-cond}
k_{00} \equiv \frac{A^2}{r_0^2}\,, \qquad \qquad
k_{01} \equiv 0\,.
\end{equation}
In the next section, on the other hand, we will not assume any particular
coordinate condition of the type (\ref{coor-cond}), leaving a possibility
to fix it in any alternative way.

\subsection{Asymptotic properties}
\label{sec:asymptotic}

Following the discussion on the asymptotic properties of the parametrized
metric made with expression (\ref{BL}), we can deduce the following
asymptotic behavior of the line element \eqref{fixedmetric}
\begin{subequations}
\label{PPN}
\begin{align}
& \hskip -1.5cm
\frac{N^2(r,\theta)-W^2(r,\theta)\sin^2\theta}{K^2(r,\theta)} =
\nonumber\\
&= \frac{N^2(r,\theta)}{K^2(r,\theta)-{AW(r,\theta)}/{r}}+{\cal
  O}\left(\frac{1}{r^3}\right)\nonumber\\
&= 1-\frac{2M}{r}+(\beta-\gamma)\frac{2M}{r^2} + {\cal
  O}\left(\frac{1}{r^3}\right)\,,\\ \nonumber\\
W(r,\theta)
&= \frac{2J}{r^2}+{\cal O}\left(\frac{1}{r^3}\right) =
  \frac{2MA}{r^2}+{\cal O}\left(\frac{1}{r^3}\right)\,,\\
\nonumber\\
\frac{B^2(r,\theta)}{N^2(r,\theta)}
&= 1+\gamma\frac{2M}{r} + {\cal
    O}\left(\frac{1}{r^2}\right)\,,\\
\nonumber\\
K^2(r,\theta)
&= 1+{\cal O}\left(\frac{1}{r^2}\right)\,.
\end{align}
\end{subequations}
Note that in deriving expressions \eqref{PPN} we have also introduced the
PPN parameters $\beta$ and $\gamma$ \cite{Will:2005va} and assumed a
reflection symmetry across the equatorial plane, \ie the functions in
(\ref{yexp}) are taken to depend only on $y^2$ ($i =
0,2,4\ldots$). Furthermore, adopting the classification made by Thorne in
Ref. \cite{Thorne:1980ru}, metrics of this type are referred to as
``Cartesian and mass centered to order 0'' (ACMC-0). Clearly, using
expressions (\ref{PPN}), it is possible to read off the mass $M$ and
angular momentum $J$ of the spacetime. We note that although the
asymptotic behavior we have chosen is well motivated from an
astrophysical point of view, our approach is not limited by any
particular choice of asymptotic behavior and can, in principle, be
constructed also for axisymmetric black holes having different asymptotic
constraints.

Next, for a metric with the asymptotic behavior given by expressions
(\ref{PPN}), we find that
\begin{subequations}
\label{asympcoeff}
\begin{eqnarray}\label{epsilon0}
\epsilon_0& = &\frac{2M-r_0}{r_0}\,,\\
\label{a00}
a_{00}& = &(\beta-\gamma)\frac{2M^2}{r_0^2} =
\frac{(\beta-\gamma)(1+\epsilon_0)^2}{2}\,,\\
\label{b00}
b_{00}& = &(\gamma-1)\frac{M}{r_0} = \frac{(\gamma-1)(1+\epsilon_0)}{2}\,,\\
\label{w00}
w_{00}& = &\frac{2J}{r_0^2} = \frac{J}{M^2}\frac{(1+\epsilon_0)^2}{2}\,,\\
\label{otherasympcoeff}
\epsilon_i& = &0 = b_{i0} = w_{i0}\,,\quad i>0.
\end{eqnarray}
\end{subequations}
The asymptotic parameters, $a_{i0}$ and $k_{i0}$ for $i>0$, are not fixed
only by $M$, $J$, $\beta$, and $\gamma$. In particular, the parameter
$a_{20}$ contains also information on the quadrupole moment of the black
hole, which cannot be read off from the asymptotic expansion. This is
because the metric with components (\ref{fdef}) and with the parameters
fixed by expressions (\ref{asympcoeff}), is not of type ACMC-1 unless
$b_{i1} = 0$ and $k_{i0} = 0$ for all $i>0$. In this latter case,
$a_{20}$ is related to the multipole $I^{20}$ introduced by Thorne in
\cite{Thorne:1980ru}; in particular, using expression (11.4a) of
\cite{Thorne:1980ru}, we can read off the value of $a_{20}$ as
\begin{equation}
I^{20} = -\frac{4}{3}\sqrt{\frac{4\pi}{15}}{a_{20}r_0^3}\,.
\end{equation}

Note that since $b_{i1}$ are parameters fixed near the black-hole
horizon, we are unable to find a general transformation from
(\ref{fixedmetric}) to an ACMC-1 type of metric and cannot, therefore,
express $I^{20}$ only in terms of the asymptotic
parameters. Nevertheless, as we will show in the next section, $a_{i0}$
and $k_{i0}$ can be related to observable quantities in a way similar to
what was done for the Geroch-Hansen quadrupole moment of the black hole
\cite{GerochHansen}. In particular, we will compare the orbital-plane
precession frequency with the formula derived by Ryan \cite{Ryan:1995wh},
thus obtaining a definition of the quadrupole moment through the
asymptotic parameters only.

\section{Circular geodesic in the equatorial plane and multipole moments}
\label{sec:circulargeo}

Following \cite{Ryan:1995wh}, we consider a circular geodesic motion in
the equatorial plane, \ie with
\begin{equation}
\frac{dr}{dt} = 0\,,\qquad \theta = \frac{\pi}{2}\,,
\end{equation}
and with orbital frequency
\begin{equation}\label{Omega}
\Omega\equiv\frac{d\phi}{dt} = \frac{-g_{t\phi,r}+\sqrt{g_{t\phi,r}^2 -
g_{tt,r}g_{\phi\phi,r}}}{g_{\phi\phi,r}}\,,
\end{equation}
where we use a comma to indicate a partial derivative. The energy per
unit mass is then given by (see, \eg \cite{RezzollaZanotti})
\begin{equation}\label{Em}
\frac{E}{m} = \frac{-g_{tt}-g_{t\phi}\Omega}{\sqrt{-g_{tt} -
2g_{t\phi}\Omega-g_{\phi\phi}\Omega^2}}\,,
\end{equation}
and, after using expression \eqref{Omega}, it can be expressed as a
function of $\Omega$ only. After replacing in (\ref{Em}) the
parametrized expressions for the metric functions, we can derive an
expression for the energy per logarithmic interval of frequency in terms
of the dimensionless variable $v\equiv (\Omega M)^{1/3}$
\begin{widetext}
\begin{eqnarray}
\label{dE}
\frac{\Delta E}{m} &=& -\frac{1}{m}\frac{dE}{d\log{\Omega}} = -
\frac{v}{3m}\frac{dE}{dv}
= \frac{v^2}{3}+v^4\left(-\frac{1}{2}+\frac{8a_{00}}{9(1+\epsilon_0)^2}\right)
+ v^5\frac{40w_{00}}{9(1+\epsilon_0)^2} + \\\nonumber
&\phantom{=}& \hskip 1.0cm
v^6\left[-\frac{27}{8}+\frac{6a_{00}+4k_{00}-4w_{00}A/M}{(1+\epsilon_0)^2}
+ \frac{16a_{00}^2}{3(1+\epsilon_0)^4} +
8\frac{\epsilon_0-a_{00}-k_{00}+{\tilde A}_0(1)-{\tilde
    K}_0(1)}{(1+\epsilon_0)^3}\right]+{\cal O}(v^7)\,,
\end{eqnarray}
\end{widetext}
where
\begin{align}
{\tilde A}_0(1) &=
\dfrac{a_{01}}{1+\dfrac{a_{02}}{1+\dfrac{a_{03}}{1+\ldots}}}\,, \\
\nonumber \\
{\tilde K}_0(1) &= \dfrac{k_{01}}{1+\dfrac{k_{02}}{1+\dfrac{k_{03}}{1+\ldots}}}\,.
\end{align}

In a similar way, we can obtain a series expansion of the periastron
precession frequency $\Omega_r$ and orbital-plane precession frequency
$\Omega_\theta$, defined as \cite{Ryan:1995wh}
\begin{eqnarray}
\Omega_r& \equiv &\Omega-\sqrt{-\frac{A_{_E}^2G_{\phi\phi,rr}+2A_{_E}A_{_L}G_{t\phi,rr}
    + A_{_L}^2G_{tt,rr}}{2g_{rr}}}\,,\nonumber\\ \\
\Omega_\theta& \equiv &\Omega-\sqrt{-\frac{A_{_E}^2G_{\phi\phi,\theta\theta} +
2A_{_E}A_{_L}G_{t\phi,\theta\theta}+A_{_L}^2G_{tt,\theta\theta}}{2g_{\theta\theta}}}\,,
\nonumber\\
\end{eqnarray}
where
\begin{eqnarray}
A_{_E}& \equiv &-g_{tt}-g_{t\phi}\Omega =
\frac{N^2}{K^2}+W\sin^2\theta\left(r\Omega-\dfrac{W}{K^2}\right)\,,\nonumber\\ \\
A_{_L}& \equiv &g_{t\phi}+g_{\phi\phi}\Omega = rK^2\sin^2\theta\left(r\Omega -
\dfrac{W}{K^2}\right)\,, \\
G_{\alpha\beta}& \equiv
&\frac{g_{\alpha\beta}}{g_{t\phi}^2-g_{tt}g_{\phi\phi}} =
\frac{g_{\alpha\beta}}{N^2r^2\sin^2\theta}\,,\quad\alpha,\beta = t,\phi\,.
\end{eqnarray}

Collecting terms, we can express the normalized periastron precession and
orbital-plane precession frequencies respectively as
\begin{widetext}
\begin{eqnarray}
\label{Ot}
\frac{\Omega_\theta}{\Omega} &=& -v^2\frac{2\epsilon_2}{(\epsilon_0+1)^2}
  + v^3\frac{4(w_{00}-w_{20})}{(\epsilon_0+1)^2} +
  v^4\left[-\frac{2(k_{20}+k_{00}-w_{20}A/M)}{(\epsilon_0+1)^2} -
  \frac{4a_{20}}{(\epsilon_0+1)^3}\right.
  \left.+\frac{6\epsilon_2^2-32\epsilon_2a_{00}}{3(\epsilon_0+1)^4}\right]+{\cal
    O}(v^5)\,,\nonumber\\ \\
\frac{\Omega_r}{\Omega} &=&
  v^2\left[3-\frac{2a_{00}}{(\epsilon_0+1)^2}+\frac{b_{00}}{2(\epsilon_0+1)}\right]
  - \frac{8w_{00}v^3}{(\epsilon_0+1)^2}
  +v^4\left[\frac{9}{2}-\frac{3b_{00}}{2(\epsilon_0+1)} -
  \frac{8a_{00}+6k_{00}-6w_{00}A/M}{(\epsilon_0+1)^2}\right.\nonumber\\
\label{Or}
&\phantom{=}&-\frac{5b_{00}^2}{8(\epsilon_0+1)^2}+\frac{5b_{00}a_{00}}{3(\epsilon_0+1)^3}-
\frac{26a_{00}^2}{3(\epsilon_0+1)^4}
\left.+\frac{12(a_{00}+k_{00}+k_{01}-\epsilon_0-{\tilde
    A}_0(1))}{(\epsilon_0+1)^3}+\frac{{\tilde
    B}_0(1)}{(\epsilon_0+1)^2}\right]+{\cal O}(v^5)\,,
\end{eqnarray}
\end{widetext}
where
\begin{equation}
{\tilde B}_0(1) = \dfrac{b_{01}}{1+\dfrac{b_{02}}{1+\dfrac{b_{03}}{1+\ldots}}}\,.
\end{equation}

We can now compare expressions \eqref{Ot} and \eqref{Or} with the
corresponding ones obtained by Ryan in general relativity
\cite{Ryan:1995wh}, \ie
\begin{subequations}
\label{Ryan}
\begin{eqnarray}
\frac{\Delta E}{m}& =
&\frac{v^2}{3}-\frac{v^4}{2}+v^5\frac{20S_1}{9M_0^2}+v^6\left(-\frac{27}{8}+
\frac{M_2}{M_0^3}\right)+{\cal O}(v^7)\,, \nonumber\\ \\
\frac{\Omega_\theta}{\Omega}& =
&v^3\frac{2S_1}{M_0^2}+v^4\frac{3M_2}{2M_0^3}+{\cal O}(v^5)\,,\\
\frac{\Omega_r}{\Omega}& =
&3v^2-v^3\frac{4S_1}{M_0^2}+v^4\left(\frac{9}{2}-
\frac{3M_2}{2M_0^3}\right)+{\cal O}(v^5)\,,
\end{eqnarray}
\end{subequations}
where $M_0 = M$, $S_1 = J$, and $M_2 = Q$ are the first three
Geroch-Hansen multipole moments: the ADM mass, the angular momentum and
the quadrupole moment, respectively.

Bearing in mind that
\begin{equation}
\label{momentum-coeff}
w_{00} = \dfrac{2J}{r_0}\,, \qquad\Longleftrightarrow\qquad
\dfrac{J}{M^2} =
\dfrac{2w_{00}}{(\epsilon_0+1)^2}\,,
\end{equation}
we obtain [\cf Eqs. (\ref{asympcoeff}) for $\beta = \gamma = 1$]
\begin{eqnarray}
\label{null-coeff}
&\epsilon_2 = 0 =  w_{20} = a_{00} = b_{00} \,,\\
\label{quadrupole-coeff}
&-\dfrac{3Q}{2M^3} = \dfrac{2(k_{20}+k_{00})}{(\epsilon_0+1)^2} +
\dfrac{4a_{20}}{(\epsilon_0+1)^3}\,.
\end{eqnarray}

The expression for the quadrupole moment $Q$ through the values of the
coefficients as fixed at spatial infinity can be obtained only after
measuring the orbital-plane precession frequency (\ref{Ot}). Note that in
the case in which $\epsilon_2 = 0$, such a frequency (\ref{Ot}) does not
depend on the PPN parameters that are related to the coefficient $a_{00}$
[the fifth term in Eq. \eqref{Ot} is automatically zero in this
  case]. This property suggests that the expression
(\ref{quadrupole-coeff}) derived in general relativity remains the same
also for non-Einsteinian theories. Indeed, as we will discuss in
Sec.~\ref{sec:EDGBbh}, it provides the correct answer for the
Einstein-dilaton-Gauss-Bonnet black holes. On the other hand, the
expressions for the energy per logarithmic frequency interval (\ref{dE})
and for the orbital-plane precession frequency (\ref{Or}), allow us to
obtain those higher-order PPN parameters that are encoded, within our
formalism, in the values of the coefficients ${\tilde A}_0(1)$, ${\tilde
  B}_0(1)$, and ${\tilde K}_0(1)$.

On the basis of these considerations we conclude that the quadrupole
moment of the black hole in our parametrized metric can be measured by
fitting the equatorial-plane precession frequency for a circular orbit
(\ref{Ot}). Quite generically, the latter depends on three coefficients
that are fixed asymptotically: $a_{20}$, $k_{00}$, and $k_{20}$
($w_{20}=0$ to match the results by Ryan \cite{Ryan:1995wh}). In general
relativity, similar coefficients, $M_2^*$ and $B_0$, were introduced in
Ref. \cite{Pappas:2012qg} and are related to these coefficients as
\begin{eqnarray}
&&a_{20} = -3\frac{M_2^*}{r_0^3} = -3\frac{M_2^*(\epsilon_0+1)^3}{8M^3}\,,\\
&&k_{00} = \frac{M^2+4B_0}{r_0^2} = (\epsilon_0+1)^2\left(\frac{1}{4} +
\frac{B_0}{M^2}\right)\,,\\
&&k_{20} = 0\,.
\end{eqnarray}

In particular, for the Kerr metric one has that
\begin{align}
M_2^*& = -\frac{2Ma^2}{3}& &\Longleftrightarrow& a_{20} &=
\frac{2Ma^2}{r_0^3} = \frac{a^2(\epsilon_0+1)^3}{4M^2}\,,&
\\
B_0& = -\frac{M^2-a^2}{4}& &\Longleftrightarrow& k_{00} &=
\frac{a^2}{r_0^2} = \frac{a^2(\epsilon_0+1)^2}{4M^2}\,.&
\end{align}
Note that unlike in \cite{Pappas:2012qg}, the value of $k_{00}$ is fixed
here by our coordinate choice. One could choose coordinates such that
$k_{00}$ would assume a different value, but this would not alter the
final value of $Q$, as it is easy to verify after substituting the
corresponding values of $a_{20}$ and $k_{20}$ in
(\ref{quadrupole-coeff}). As a result, with our choice of coordinates
leading to expressions (\ref{coor-cond}), we will always have $k_{00} =
a^2/r_0^2$, so that the quadrupole moment is completely determined by the
coefficient $a_{20}$.

\section{Parametrized representation for the rotating dilaton black hole}
\label{sec:dilbh}

This section is dedicated to the explicit calculation of the coefficients
needed for the representation of the parametrized metric that
approximate the rotating dilaton black-hole spacetime
(\ref{dilatonmetric}) \cite{dilaton-BH}. This is an important test of our
approach and an example of a metric that is more complex than the Kerr
solution.

We first substitute (\ref{dilatonmetric_3})--(\ref{dilatonmetric_5}) in
(\ref{alpha0}) and taking into account that $A = a$ for a dilaton black
hole, we find that
\begin{equation}
\alpha_0(\rho) = 0\,,
\end{equation}
while from expressions (\ref{coordtransform}) obtain the relations
\begin{subequations}
\label{yexpdilaton}
\begin{eqnarray}
y &=& \vary\,, \\
r^2 &=& \rho(\rho+2b)\,,\\
N^2 &=& \frac{(\sqrt{b^2+r^2}-b)^2-2\mu(\sqrt{b^2+r^2}-b)+a^2}{r^2}\,,\\
W &=& \frac{2a(\mu+b)(\sqrt{r^2+b^2}-b)}{r(r^2+a^2y^2)}\,,\\
B^2 &=& \left(\frac{d\rho}{dr}\right)^2 = \frac{r^2}{b^2+r^2}\,,\\
K^2 &=& \left(1+\frac{a^2y^2}{r^2}\right)^{-1}
\left[\left(1+\frac{a^2}{r^2}\right)^2-\frac{a^2(1-y^2)}{r^2}N^2\right]
\,. \nonumber \\
\end{eqnarray}
\end{subequations}
These expressions coincide up to ${\cal O}(a^2)$ with the slowly rotating
metric already described in \cite{RezzollaZhidenko}. Furthermore, since
$N$ does not depend on $y$, the relation
\begin{equation}
r_0 = \sqrt{(\mu+b+\sqrt{\mu^2-a^2})^2-b^2} \,,
\end{equation}
defines the event horizon and reduces to expression (55) of
\cite{RezzollaZhidenko} when terms of ${\cal O}(a^2)$ are neglected.

By comparing the series expansion of (\ref{yexp}) and (\ref{yexpdilaton})
near spatial infinity (\ie $x = 1$) we find that
\begin{subequations}\label{dilatonascoeff}
\begin{eqnarray}
\epsilon_0& = &\frac{2b+2\mu-r_0}{r_0}\,,\\
a_{00}& = &\frac{2b(b+\mu)}{r_0^2}\,,\\
b_{00}& = &0\,,\\
w_{00}& = &\frac{2a(b+\mu)}{r_0^2}\,,
\end{eqnarray}
which, by comparison with (\ref{asympcoeff}), give correct values for the
ADM mass $M$ and the angular momentum $J$, namely
\begin{equation}
M = \mu+b\,,\qquad\qquad J = Ma\,,
\end{equation}
and reproduce the same PPN coefficients as for the nonrotating case
\cite{RezzollaZhidenko},
\begin{equation}
\beta = \frac{\mu+2b}{\mu+b} = 1+\frac{b}{M}\,,
\qquad\qquad\gamma = 1\,.
\end{equation}
We also we find that
\begin{equation}
a_{20} = \frac{2a^2(\mu+b)}{r_0^3}= \frac{2Ma^2}{r_0^3}\,,
\end{equation}
while all the other asymptotic parameters are zero, except for $k_{00}$,
which is given by
\begin{equation}
k_{00} = \frac{a^2}{r_0^2}\,.
\end{equation}
\end{subequations}

From expression (\ref{quadrupole-coeff}), we find the expected result
that the quadrupole moment is given by
\begin{equation}\label{quadrupole-moment}
Q = -\frac{a_{20}r_0^3+M (k_{00}+k_{20})r_0^2}{3} = -Ma^2 =
-\frac{J^2}{M}\,,
\end{equation}
for any value of $b$.

Finally, by comparing the series expansion of (\ref{yexp}) and
(\ref{yexpdilaton}) near $x = 0$ we find that
\begin{subequations}\label{dilatonhorcoeff}
\begin{eqnarray}
a_{01} &=& \frac{2(\mu+b)(2b^2+r_0^2+(2r_0-3b)\sqrt{r_0^2+b^2})}
{r_0^2\sqrt{r_0^2+b^2}}\nonumber \\
&&-3\frac{r_0^2+a^2}{r_0^2}\,, \\
a_{21} &= &-\frac{a^4+2a^2(\mu+b)
(b+r_0-\sqrt{r_0^2+b^2})}{r_0^4}\,,\\
a_{11} &=& 0 = a_{31} = a_{41} = a_{51}\,,\ldots\,,\\
b_{01} &=& \frac{r_0}{\sqrt{r_0^2+b^2}}-1\,,\\
b_{11} &=& 0 = b_{21} = b_{31} = b_{41} = b_{51}\,,\ldots\,,\\
w_{01} &=& \frac{2a(\mu+b)(b+r_0-\sqrt{r_0^2+b^2})}{r_0^3}\,,\\
w_{11} &=& 0 = w_{21} = w_{31} = w_{41} = w_{51}\,,\ldots\,, \\
k_{21} &=& \frac{a^4-2a^2(\mu+b)(\sqrt{r_0^2+b^2}-b)}{r_0^4}\,,\\
k_{01} &=& 0 = k_{11} = k_{31} = k_{41} = k_{51}\,,\ldots\,,
\end{eqnarray}
\end{subequations}
and so on. Clearly, $a_{01}$, $b_{01}$, and $w_{01}$ coincide with the
corresponding parameters of the slowly rotating dilaton black hole.

As a final remark, we reinforce a statement already made in
Sec. \ref{sec:parametrization}.  While the line element of the Kerr
spacetime can be reproduced exactly within the proposed parametrization
with a finite number of terms of the Pad\'e expansion, this does not seem
to be possible for the dilaton black hole, whose metric functions are not
a ratio of polynomials in powers of $r$, but rather contain square roots
of polynomials.

\subsection{Testing the parametrization in the equatorial plane:
The binding energy}

Although the parametrization suggested here can be used for generic
investigations of the physics of black holes, our main interest is to
find a general parametrization for a rotating black-hole spacetime which
allows us to model processes connected to electromagnetic emission from
matter accreting onto the supermassive black-hole candidate at the center
of our Galaxy \cite{BlackHoleCam}. Because of this, it is important to
test the ability of our parametrization in describing accurately
radiative processes near the event horizon and, for instance, the
formation of a shadow. This will be the focus of a forthcoming companion
paper \cite{Younsietal}, but some preliminary estimate can already be
presented here in terms of the calculation of the simplest basic quantity
entering in the physics of accretion flows onto black holes: the
\emph{binding energy} of a particle moving on a circular orbit.

\begin{table*}
\begin{tabular}{|c|c|c|c|c|}
\hline
$b$ & $1\rm{st\ order}$ & $2{\rm nd\ order}$ & $3{\rm rd\ order}$ & ${\rm exact}$ \\
\hline
0.00& 5.719095841793664& 5.719095841793664& 5.719095841793664& 5.719095841793664\\
0.02& 5.771085348147105& 5.771085348949989& 5.771069849415838& 5.771085349099403\\
0.20& 6.207254764244374& 6.207261486774459& 6.207262811802919& 6.207262810978281\\
0.50& 6.832236606077545& 6.832430295600210& 6.832473869879982& 6.832473613387891\\
1.00& 7.676903064196137& 7.678772618393546& 7.679324761207151& 7.679311080887710\\
\hline
\end{tabular}
\caption{Binding energies as computed at different orders in the radial
  (continued-fraction) expansion for a nonrotating black hole with a
  dilaton field $b$. The last column refers to the exact
  solution.}\label{tabl:a = 0}
\end{table*}

To this scope, we calculate the energy of the particle at the innermost
stable circular orbit (ISCO) and thus the binding energy as the amount of
energy released by the particle going over from a given stable circular
orbit located at $r_0$ over to the ISCO at $r_{_{\rm ISCO}}$, \ie
\begin{equation}
\text{BE} = 1 - \frac{E(r_{_{\rm ISCO}})}{E(r_0)}\,,
\label{BE}
\end{equation}
where the initial circular orbit $r_0$ is normally considered to be at
spatial infinity but needs not be. The binding energy of massive test
particles is rather sensitive to the black-hole properties and in
Ref. \cite{Konoplya:2006qr} it was calculated for a number of deformed
black-hole spacetimes. We here recall that in the case of an extremal
Kerr black hole it equals $\simeq 3.8\%$ for counterrotating particles
and $\simeq 42\%$ for corotating ones, while it is $\simeq 5.7\%$ for a
Schwarzschild black hole. The binding energy also increases above $40\%$,
when the dilaton $b$ ranges from $0$ to $1$.

Defining the four-momentum of a massive particle as
\begin{equation}
p^\alpha \equiv m \frac{dx^\alpha}{d s}\,,
\end{equation}
where $s$ is an invariant affine parameter, we also recall that in a
stationary, axisymmetric metric there are three integrals of motion which
can be related to the particle's rest mass $m$, to its energy $E = -p_t$,
and its angular momentum $L = p_\phi$. The normalization condition on the
four-momentum
\begin{equation}
p_\alpha p^\alpha = -m^2
\label{equation1}
\end{equation}
leads to the following relation in the equatorial plane
\begin{equation}
m^2 g_{rr}\left(\frac{dr}{d s}\right)^2 = V_{\rm eff}(r)\,,
\end{equation}
where the effective potential is defined as
\begin{eqnarray}
\label{effective-potential}
&&V_{\rm eff}(r) \equiv - \left( g^{tt}E^2 - 2g^{t\phi}E L + g^{\phi\phi}L^2 + m^2
  \right)\Biggr|_{\theta = \pi/2}\!\!\!\!\! =  \\\nonumber
&&\frac{K^2(r,\pi/2)}{N^2(r,\pi/2)}
\left(E-\frac{W(r,\pi/2)}{K^2(r,\pi/2)}\frac{L}{r}\right)^2\!\!-\frac{L^2}{r^2
    K^2(r,\pi/2)} - m^2\,.
\end{eqnarray}
The energy $E$ and momentum $L$ of a particle on a circular orbit at
radial position $r$ can then be determined from the requirements that
\begin{equation}
V_{\rm eff}(r)  = 0, \qquad V_{\rm eff}'(r) = 0\,,
\end{equation}
where $'$ indicates a derivative in the radial direction. Once the
expressions for $L(r)$ and $E(r)$ have been obtained in this way, the
position of the ISCO is computed from the additional condition $V_{\rm
  eff}''(r) = 0$ and then to compute the value of the energy at the ISCO
$E(r_{_{\rm ISCO}})$, and hence the binding energy as in expression
(\ref{BE})\footnote{We recall that, as already noted in
  \cite{RezzollaZhidenko} in the slowly rotating regime, the position of
  the ISCO does not depend on the function $B$. }.

\begin{table*}
\begin{tabular}{|c|c|r|r|r|r|r||r|r|r|r|r|}
\hline
$a/\mu$ & $b$ &
$1\rm{st\ order}$ & $2{\rm nd\ order}$ & $3{\rm rd\ order}$ & $4{\rm th\ order}$ & ${\rm exact}$ &
$1\rm{st\ order}$ & $2{\rm nd\ order}$ & $3{\rm rd\ order}$ & $4{\rm th\ order}$ & ${\rm exact}$ \\
\hline
0.50 & 0.00 & 8.2118 & 8.2118 & 8.2118 & 8.2118 & 8.2118 & 4.5142 & 4.5142 & 4.5142 & 4.5142 & 4.5142 \\
0.50 & 0.02 & 8.2676 & 8.2675 & 8.2675 & 8.2675 & 8.2675 & 4.5611 & 4.5611 & 4.5611 & 4.5611 & 4.5611 \\
0.50 & 0.20 & 8.7368 & 8.7317 & 8.7318 & 8.7318 & 8.7318 & 4.9555 & 4.9568 & 4.9568 & 4.9568 & 4.9568 \\
0.50 & 0.50 & 9.4142 & 9.3873 & 9.3881 & 9.3881 & 9.3881 & 5.5239 & 5.5315 & 5.5311 & 5.5311 & 5.5311 \\
0.50 & 1.00 & 10.3402& 10.2568& 10.2610& 10.2612& 10.2613& 6.2972 & 6.3230 & 6.3217 & 6.3214 & 6.3214 \\
\hline
\hline
0.95 & 0.00 & 19.0144 & 19.0144 & 19.0144 & 19.0144 & 19.0144 & 3.8362 & 3.8362 & 3.8362 & 3.8362 & 3.8362 \\
0.95 & 0.02 & 19.0083 & 19.0084 & 19.0084 & 19.0084 & 19.0084 & 3.8791 & 3.8791 & 3.8791 & 3.8791 & 3.8791 \\
0.95 & 0.20 & 19.0140 & 19.0115 & 19.0098 & 19.0097 & 19.0097 & 4.2405 & 4.2430 & 4.2429 & 4.2428 & 4.2428 \\
0.95 & 0.50 & 19.2262 & 19.1424 & 19.1284 & 19.1269 & 19.1270 & 4.7629 & 4.7767 & 4.7757 & 4.7755 & 4.7755 \\
0.95 & 1.00 & 19.9278 & 19.4719 & 19.4191 & 19.4121 & 19.4133 & 5.4763 & 5.5219 & 5.5184 & 5.5170 & 5.5171 \\
\hline
\end{tabular}
\caption{The same as in Table \ref{tabl:a = 0}, but for a rotating
  dilaton black hole with $a = 0.5\mu$ (upper rows) and $a = 0.95\mu$
  (lower rows). The left columns refer to corotating particles, while
  the right ones to counterrotating particles. For all the cases
  considered, the polar expansion was kept at the order in $\cos^2\theta$
  (\ie at $\cos^4\theta)$, since higher orders are zero for a dilaton
  black hole.}\label{tabl:a = 0.5}
\end{table*}

Table \ref{tabl:a = 0} reports the binding energies as computed at
different orders in the radial (continued-fraction) expansion for a
nonrotating black hole with a dilaton field $b$. The last column refers
to the exact solution, \ie to the metric \eqref{dilatonmetric}. Clearly,
the convergence in the spherically symmetric case is excellent and the
first order is sufficient to obtain a relative error which is $\sim
10^{-4}$ in the most extreme case of $b=1$. This relative error in the
binding energy further reduces to $\sim 10^{-6}$ when considering the
expansion at fourth order; clearly these uncertainties are far smaller
than what is to be expected from astronomical observations.

The convergent behavior is only slightly less good when rapid rotation
is introduced. This is collected in Table \ref{tabl:a = 0.5}, which
refers to a rotating dilaton black hole with $a = 0.5\mu$ (upper rows)
and $a = 0.95\mu$ (lower rows), and where the left columns refer to
corotating particles, while the right ones to counterrotating
particles. In particular, at second order the errors are about $\sim
10^{-4}$ and $\sim 10^{-3}$ and for $a = 0.5\mu, b=1$ and $a = 0.95\mu$,
$b=1$, respectively. We note that as the binding energy of a
quasiextreme rotating dilaton black hole reaches its maximum of $42\%$
for $a \simeq 0.99999 \mu$, reproducing this limiting value with the
parametrization would apparently require an expansion up to very high
orders in the radial direction. However, since in the equatorial plane
the Kerr spacetime (\ie $b=0$) is reproduced exactly already at the first
order of the continued fraction, the value of binding energy coincides
with the exact binding energy.

\section{Parametrized representation for rotating EDGB black holes}
\label{sec:EDGBbh}

While a rotating dilaton black hole corresponds to an essentially
Einsteinian theory of gravity with an extra field, a theory containing
higher-curvature corrections represents a genuinely non-Einsteinian
gravitational theory. In higher than four dimensions, the second order in
curvature term (or Gauss-Bonnet term) is the dominating one.  In a
four-dimensional ($D = 4$) spacetime as the one considered here, the
Gauss-Bonnet term is invariant and leads to solutions of the Einstein
equations that are not affected unless the dilaton is coupled to the
system. Although an exact solution is unknown for such $D = 4$ rotating
dilaton-Gauss-Bonnet black holes, an approximate metric has been deduced
in the regime of slow rotation \cite{Ayzenberg:2014aka}.

In this case the solution has two small parameters
\begin{equation}
\chi\equiv \frac{a}{M} = \frac{J}{M^2}\,, \qquad \qquad
\zeta\equiv\frac{16\pi \alpha^2}{\beta M^4}\,,
\end{equation}
where $\alpha$ and $\beta$ are the two coupling constants
Einstein-dilaton-Gauss-Bonnet theory, with the first one being related to
the coupling of higher curvature, while the second one accounts for the
coupling with the scalar field. After some algebra, the metric functions
up to the order ${\cal O}(\chi^3,\zeta^2)$ are given by
\cite{Ayzenberg:2014aka}
\begin{widetext}
\begin{eqnarray}
f(\rho,\vartheta) &=& 1 - \frac{2 M}{\rho } +\frac{M^2 \chi^2}{\rho^2}
  +\zeta\left( - \frac{80 M^7}{3 \rho^7} +\frac{32 M^6}{5 \rho^6}
  +\frac{22 M^5}{5 \rho^5} +\frac{26 M^4}{3 \rho^4} +\frac{M^3}{3 \rho
   ^3}\right)+\\
&\phantom{=}& \chi^2\zeta\Biggr[ - \frac{80 M^{11}}{\rho^{11}} +\frac{7336
  M^{10}}{45 \rho^{10}} - \frac{20422 M^9}{315 \rho^9} +\frac{1917
  M^8}{245 \rho^8} - \nonumber\\
&\phantom{=}& \hskip 2.8cm \frac{253756 M^7}{11025 \rho^7} +\frac{838039
  M^6}{110250 \rho^6} - \frac{18551 M^5}{5250 \rho^5} - \frac{3048
  M^4}{875 \rho^4} - \frac{M^3}{6 \rho^3}+ \nonumber \\
&\phantom{=}& \hskip -0.2cm \cos^2\vartheta\Biggr(\frac{240 M^{11}}{\rho^{11}} - \frac{6136
  M^{10}}{15 \rho^{10}} +\frac{9214 M^9}{21 \rho^9} - \frac{26233
  M^8}{245 \rho^8} +\frac{30316 M^7}{3675 \rho^7} - \frac{1497089
  M^6}{36750 \rho^6} +\frac{11201 M^5}{1750 \rho^5} + \frac{3019
  M^4}{875 \rho^4}\Biggr)\Biggr]\,,\nonumber
\end{eqnarray}

\begin{eqnarray}
\beta^2(\rho,\vartheta) &=& 1 - \zeta\left(\frac{48 M^6}{\rho^6} +
  \frac{128 M^5}{5 \rho^5} + \frac{14 M^4}{\rho^4} + \frac{8 M^3}{3
    \rho^3} + \frac{M^2}{\rho^2}\right) +\\
&\phantom{=}& \chi^2\zeta\Biggr[ - \frac{720 M^{10}}{\rho^{10}} +
    \frac{28688 M^9}{45 \rho^9} + \frac{2102 M^8}{15 \rho^8} + \frac{616
      M^7}{5 \rho^7} - \frac{907 M^6}{45 \rho^6} + \frac{139 M^5}{15
      \rho^5} + \frac{5 M^4}{\rho^4} + \frac{4 M^3}{3 \rho^3} +
    \frac{M^2}{2 \rho^2} + \nonumber\\
&\phantom{=}& \cos^2\vartheta\Biggr(\frac{2160 M^{10}}{\rho^{10}} -
  \frac{26288 M^9}{15 \rho^9} + \frac{1658 M^8}{5 \rho^8} - \frac{208
    M^7}{5 \rho^7} + \frac{2767 M^6}{15 \rho^6} + \frac{11 M^5}{\rho^5} +
  \frac{2 M^4}{\rho^4}\Biggr)\Biggr]\,,\nonumber
\end{eqnarray}
\begin{eqnarray}
\sigma^2(\rho,\vartheta)& = &1 + \frac{M^2 \chi^2
  \cos^2\vartheta}{\rho^2} - \chi^2\zeta(3\cos^2\vartheta - 1)
Q(\rho)\,,\\
\omega(\rho,\vartheta)& = &\frac{2 M^2 \chi }{\rho^2}\Biggr[1 +
\zeta\Biggr(\frac{40 M^6}{3 \rho^6} - \frac{24 M^5}{5 \rho^5} - \frac{3
  M^4}{\rho^4} - \frac{14 M^3}{3 \rho^3} - \frac{3 M^2}{10
  \rho^2}\Biggr)\Biggr]\,,\\
\kappa^2(\rho,\vartheta)& = &1 + \frac{M^2 \chi^2}{\rho^2} + \frac{2 M^3
  \chi^2\sin^2\vartheta}{\rho^3} - \chi^2\zeta(3\cos^2\vartheta - 1)
Q(\rho)\,,
\end{eqnarray}
where
\begin{equation}
Q(\rho) \equiv \frac{4463 M^3}{2625 \rho^3}+\frac{2074 M^4}{525
  \rho^4}+\frac{266911 M^5}{36750 \rho^5}+\frac{12673 M^6}{1575 \rho^6} -
\frac{12371 M^7}{735 \rho^7} - \frac{3254 M^8}{105 \rho^8} - \frac{2536
  M^9}{45 \rho^9}+\frac{80 M^{10}}{\rho^{10}}\,.
\end{equation}
\end{widetext}

The event horizon in these coordinates is a sphere with radius
\begin{equation}
\rho_0 = 2M \left(1-\frac{\chi^2}{4}-\frac{49 \zeta }{80}-
\frac{277\chi^2\zeta}{1920}\right)+{\cal O}(\chi^4,\zeta^2)\,.
\end{equation}

We note that the new coordinates ($A = M\chi$) are
\begin{equation}\nonumber
y = \vary+{\cal O}(\chi^2\zeta)\,,\qquad r = \rho+{\cal O}(\chi^2\zeta)\,,
\end{equation}
where the correction $\propto\chi^2\zeta$ can be found as a series
expansion with respect to $\vary$. Moreover, we have found that
$\forall\rho>2M$, and the following conditions are satisfied
\begin{equation}
\label{convergenceEDGB}
|\alpha_8(\rho)|>|\alpha_6(\rho)|>|\alpha_4(\rho)|
\end{equation}
and
\begin{equation}
|\zeta_8(\rho)|>|\zeta_6(\rho)|>|\zeta_4(\rho)|\,,
\end{equation}
which can be interpreted as a signal of the absolute convergence of the
series for all $|\vary|\leq1$.

\begin{widetext}
The relation between the coordinates $(y,r)$ and $(\vary,\rho)$ is given
by
\begin{eqnarray}
\label{coords_EDGB_1}
y &=& \vary \Biggl[1+
  \chi^2\zeta\Biggr(\frac{40 M^9}{3 \rho^9}-\frac{24
  M^8}{5 \rho^8}-\frac{3 M^7}{\rho^7}-\frac{14 M^6}{3 \rho^6}-\frac{3
  M^5}{10 \rho^5}+{\cal O}(\vary^2)\Biggr)\Biggr]\,,\\
\label{coords_EDGB_2}
r^2 &=& \rho\Biggl[1 \!+\!\chi^2\zeta\Biggr(\frac{80
  M^{10}}{\rho^{10}}-\frac{3736 M^9}{45 \rho^9}-\frac{2246 M^8}{105
  \rho^8}-\frac{7961 M^7}{735\rho^7} \!+\!\frac{27373 M^6}{1575
  \rho^6} \!+\!\frac{288961 M^5}{36750 \rho^5}  \!+\!\frac{2074 M^4}{525
  \rho^4} \!+\!\frac{4463 M^3}{2625 \rho^3} \!+\!
  {\cal O}(\vary^2)\Biggr)\Biggr],
\nonumber \\
\end{eqnarray}
so that the location of the horizon in the equatorial plane is given by
\begin{equation}
r_0 = 2M\left(1-\frac{\chi^2}{4}-\frac{49 \zeta }{80}+\frac{128171 \chi
 ^2\zeta}{588000}\right)+{\cal O}(\chi^4,\zeta^2)\,.
\end{equation}

As a result of the coordinate transformations
\eqref{coords_EDGB_1} and \eqref{coords_EDGB_2}, we obtain
\begin{subequations}
\begin{eqnarray}
N^2 &=& 1 - \frac{2 M}{r} + \frac{M^2\chi^2}{r^2} +
\zeta\left(\frac{M^3}{3 r^3} + \frac{26 M^4}{3 r^4} + \frac{22 M^5}{5
  r^5} + \frac{32 M^6}{5 r^6} - \frac{80 M^7}{3 r^7}\right)\nonumber\\
&\phantom{=}& -\chi^2\zeta\Biggr(\frac{3267 M^3}{1750 r^3} +
\frac{5017M^4}{875 r^4} + \frac{136819 M^5}{18375 r^5} + \frac{35198
  M^6}{18375 r^6} - \frac{3818 M^7}{735 r^7} - \frac{4504 M^8}{245 r^8} +
\frac{16 M^9}{5 r^9}\Biggr) \nonumber \\
&\phantom{=}& +\chi^2\zeta\Biggr(\frac{3019 M^4}{875 r^4} +
\frac{6388 M^5}{875 r^5} - \frac{155394 M^6}{6125 r^6} - \frac{47878
  M^7}{1225 r^7} - \frac{17952 M^8}{245 r^8} + \frac{2040 M^9}{7 r^9} -
\frac{128 M^{10}}{5 r^{10}}\Biggr)y^2 + {\cal O}(\chi^3,\zeta^2,y^4)\,,
\nonumber \\
\end{eqnarray}
\begin{eqnarray}
B^2 &=& 1-\zeta\left(\frac{M^2}{r^2}+\frac{8 M^3}{3 r^3}+\frac{14
  M^4}{r^4}+\frac{128 M^5}{5 r^5}+\frac{48 M^6}{r^6}\right)
\nonumber\\
&\phantom{=}& +\chi^2\zeta\Biggr(\frac{M^2}{2 r^2}+\frac{4142 M^3}{875
  r^3}+\frac{2949 M^4}{175 r^4}+\frac{245724 M^5}{6125 r^5} + \frac{6028
  M^6}{105 r^6}+\frac{12792 M^7}{245 r^7}-\frac{96 M^8}{5 r^8}\Biggr)
\nonumber\\
&\phantom{=}& -\chi^2\zeta\Biggr(\frac{4463 M^3}{875 r^3}+\frac{1724
  M^4}{175 r^4}+\frac{97318 M^5}{6125 r^5}-\frac{8924 M^6}{175 r^6}-
\frac{43584 M^7}{245 r^7}-\frac{3056 M^8}{7 r^8}+\frac{896 M^9}{5
  r^9}\Biggr)y^2+{\cal O}(\chi^3,\zeta^2,y^4)\,,\nonumber \\
\end{eqnarray}
\begin{eqnarray}
W& = &\frac{2 M^2 \chi }{r^2}\Biggr[1-\zeta\Biggr(\frac{3
  M^2}{10 r^2}+\frac{14 M^3}{3 r^3}+\frac{3 M^4}{r^4}+\frac{24 M^5}{5
  r^5} -\frac{40 M^6}{3 r^6}\Biggr)\Biggr]+{\cal
  O}(\chi^3,\zeta^2,y^4)\,,\\
K^2 &=& 1+\frac{M^2 \chi^2}{r^2}+W\frac{M\chi}{r}-\frac{2M^3
  \chi^2}{r^3}y^2
\nonumber\\
&\phantom{=}& -\chi^2\zeta\Biggr(\frac{4463 M^3}{875 r^3} + \frac{2074
  M^4}{175 r^4} + \frac{127943 M^5}{6125 r^5} + \frac{4448 M^6}{525 r^6}
-\frac{2326 M^7}{245 r^7}-\frac{2792 M^8}{35 r^8} + \frac{16 M^9}{15
  r^9}\Biggr) y^2 + {\cal O}(\chi^4,\zeta^2,y^4)\,. \nonumber\\
\end{eqnarray}
\end{subequations}
\end{widetext}

Also in this case, by comparing with the series expansion (\ref{yexp})
near spatial infinity (\ie $x = 1$) we find that
\begin{subequations}\label{EDGBascoeff}
\begin{eqnarray}
\epsilon_0 &=& \frac{2M-r_0}{r_0}\,,\\
a_{00} &=& 0 = b_{00}\,,\\
w_{00} &=& \frac{2M^2\chi}{r_0^2}\,,\\
k_{00} &=& \frac{M^2\chi^2}{r_0^2}\,,
\end{eqnarray}
thus implying that the mass, angular momentum, and PPN coefficients of an
EDGB black hole obey the same relations as the ones for a Kerr black
hole. In addition, it is possible to derive that
\begin{eqnarray}
a_{20}& = &\frac{2M^3\chi^2}{r_0^3}\left(1 +
\frac{4463}{1750}\zeta\right) + {\cal O}(\chi^4,\zeta^2) \\\nonumber& =
&\frac{\chi^2}{4}\left(1 + \frac{61429 \zeta}{14000}\right) + {\cal
  O}(\chi^4,\zeta^2)\,,
\end{eqnarray}
\end{subequations}
while all of the other asymptotic parameters are zero.

The expression for the quadrupole moment can be derived from expression
(\ref{quadrupole-coeff}) and yields
\begin{eqnarray}
Q & = &-\frac{a_{20}r_0^3+M (k_{00}+k_{20})r_0^2}{3}\,,\nonumber\\
  & = &-M^3\chi^2\left(1+\frac{4463}{2625}\zeta\right)+{\cal O}(\chi^4,\zeta^2)\,,\label{EDGBquadruplole}
\end{eqnarray}
which coincides with the expression for the quadrupole moment found in
\cite{Ayzenberg:2014aka}. We should notice that our result was obtained
without the use of ACMC-1 coordinates.

Finally, by comparing the series expansions near the black-hole horizon
(\ie $x = 0$) we find that
\begin{subequations}\label{EDGBhorcoeff}
\begin{eqnarray}
a_{01} &=& -\frac{17 \zeta }{60}\left(1-\frac{324899
  \chi^2}{166600}\right)+{\cal O}(\chi^4,\zeta^2)\,,\\
b_{01} &=& -\frac{361 \zeta }{240}\left(1-\frac{51659
  \chi^2}{176890}\right)+{\cal O}(\chi^4,\zeta^2)\,,\\
w_{01} &=& -\frac{63\chi\zeta}{160}+{\cal O}(\chi^3,\zeta^2)\,,\\
k_{21} &=& -\frac{\chi^2}{4}\left(1+\frac{438867
  \zeta}{49000}\right)+{\cal O}(\chi^4,\zeta^2)\,,
\end{eqnarray}
while the other coefficients $a_{i1}$, $b_{i1}$, and $k_{i1}$ are of
order ${\cal O}(\chi^2\zeta)$, \eg
\begin{eqnarray}
a_{21} &=& \frac{447731}{392000}\chi^2\zeta+{\cal O}(\chi^4,\zeta^2)\,,\\
b_{21} &=& \frac{175629}{196000}\chi^2\zeta+{\cal O}(\chi^4,\zeta^2)\,.
\end{eqnarray}
\end{subequations}

In summary, we have shown that it is possible to obtain a full
representation of a rotating EDGB black hole within the
parametrized-metric approach introduced here. Such a representation is
specified by the metric expressions \eqref{fdef}, \eqref{tiltedfunctions}
with coefficients given by (\ref{EDGBascoeff}) and (\ref{EDGBhorcoeff}).

\section{Conclusions}
\label{sec:conclusions}

We have constructed a parametrization for a general stationary and
axisymmetric black hole which could be used for the analysis of physical
processes near rotating black holes. Our approach is based on a double
expansion in the polar and radial directions of a generic stationary and
axisymmetric metric. The polar expansion is handled via the introduction
of a series of powers of the elevation from the equatorial plane, \ie
$\cos\theta$), while the radial expansion follows the continued-fraction
approach in terms of a compactified radial coordinate that has been
successfully developed in \cite{RezzollaZhidenko} for a spherically
symmetric spacetime.

Since the parametrization uses quite general assumptions about the
spacetime of a black hole, such as the presence of Killing vectors along
the time and azimuthal coordinates, the absence of closed timelike
curves and similar pathologies of the geometry, etc., our approach is
essentially independent of any particular metric theory of gravity.

We have shown the validity and effectiveness of our approach by
reproducing accurately and with a small number of parameters three
relevant and commonly used rotating black-hole spacetimes, namely:

\begin{itemize}
\item[\emph{(i)}] a Kerr black hole, which is reproduced ``exactly'' in
  the whole space already at second order in the polar expansion [\ie at
    $\mathcal{O}(\cos^2\theta)$] and at first order in the radial
  expansion;

\item[\emph{(ii)}] a rotating dilaton black hole, which again is
  reproduced ``exactly'' at second order in the polar expansion and can
  be expanded to the desired accuracy with the expansion in the radial
  direction;

\item[\emph{(iii)}] a Gauss-Bonnet-dilaton black hole, which is
  reproduced approximately but at any desired accuracy.

\end{itemize}

The accuracy of the parametrization has been validated after comparing
the values of the binding energy for test particles moving on circular
geodesic orbits around a dilaton black hole with the exact ones. Even in
the most extreme (and realistic) cases considered, \eg for a spin of $a =
0.95\mu$ and a dilaton field $b=1$, the relative errors already at the
second order are about $\sim 10^{-3}$, and further decrease as the order
of the continued fraction is increased. Moreover, even for near-extremal
rotation, the convergence of the continued fraction is still very good,
being excellent in the equatorial plane, where it is reached already at
the few first orders of the Pad\'e approximation.

An important question which we have not addressed here, but that is
investigated in detail in a companion paper \cite{Younsietal}, is about
how many orders of the polar and radial expansions are needed for an
accurate description of physical processes outside the equatorial plane
of the black hole. Although a precise answer obviously depends on the
particular spacetime under consideration, some general statements can be
made already here. In particular, when considering the shadow cast by
various black-hole metrics, we have found that the radial expansion
through the Pad\'e approximation is always convergent. Furthermore, the
polar expansion leads to the exact solution at the second order for Kerr
and dilaton black holes, while higher-order convergence is observed for
EDGB black holes and the Johannsen-Psaltis metric. We expect therefore
that the parametrized approach presented here will be useful not only to
study generic black-hole solutions, but also to interpret the results
that will soon be made of the radio emission from the center of our
Galaxy.

\begin{acknowledgments}
It is a pleasure to thank Dimitry Ayzenberg for useful comments on
calculation of the quadrupole moment in \cite{Ayzenberg:2014aka}, as well
as Ziri Younsi, Yosuke Mizuno, Hector Olivares, and Mariafelicia de
Laurentis for numerous discussions. Partial support comes by the ERC
Synergy Grant ``BlackHoleCam - Imaging the Event Horizon of Black Holes''
(Grant No. 610058). A.~Z. was also supported by the Alexander von Humboldt
Foundation, Germany, Coordena\c{c}\~ao de Aperfei\c{c}oamento de Pessoal
de N\'ivel Superior (CAPES), Brazil, and at the final part by Conselho
Nacional de Desenvolvimento Cient\'ifico e Tecnol\'ogico (CNPq). R.~K. also
acknowledges support from the Alumni Programme of the Alexander von
Humboldt Foundation.
\end{acknowledgments}

\appendix
\section{Explicit lowest-order metric expression}
\label{sec:appendix_a}

We are aware that the derivation of the parametrized metric expressions
can appear as intricate. To facilitate the use of our parametrized
metrics, we provide here a collection of the explicit expressions of the
parametrized metrics for a Kerr, dilaton, and EDGB black hole. In each
case we do not necessarily report the highest-order expression of the
expansion. Rather, we report the orders that strike a compromise between
readability and accuracy. Hence, depending on the various cases, the
expressions reported here are either already contained in the main text
or are reported here for the first time.

We start with a brief summary of the basic expressions of the metric in
terms of the expansion coefficients and how the latter are constrained.
Hereafter we will consider the line element \eqref{fixedmetric}
\begin{widetext}
\begin{eqnarray}
\label{eq:a1}
ds^2 =
-\dfrac{N^2(r,\theta)-W^2(r,\theta)\sin^2\theta}{K^2(r,\theta)}dt^2
-2W(r,\theta)r\sin^2\theta dt \, d\phi+K^2(r,\theta)r^2\sin^2\theta d\phi^2
+\Sigma(r,\theta)\left(\dfrac{B^2(r,\theta)}{N^2(r,\theta)}dr^2 +
r^2d\theta^2\right), \nonumber \\
\end{eqnarray}
assume reflection symmetry across the equatorial plane and neglect
coefficients of higher orders. We then find
\begin{eqnarray}
\label{eq:a2}
N^2(r,\theta) &=& \left(1-\frac{r_0}{r}\right)\left[1-\frac{\epsilon_0r_0}{r}
+ \frac{(a_{00}-\epsilon_0 + k_{00})r_0^2}{r^2} +
\frac{a_{01}\,r_0^3}{r^3}\right] + \left[\frac{(k_{20} +
  \epsilon_2)r_0^2}{r^2} + \frac{(k_{21} + a_{20})r_0^3}{r^3} +
\frac{a_{21}\,r_0^4}{r^4}\right]\cos^2\theta\,,\nonumber\\\\
\label{eq:a3}
B(r,\theta) &=& 1 + \frac{b_{00}\,r_0}{r} + \frac{b_{01}\,r_0^2}{r^2} +
\left(\frac{b_{20}\,r_0}{r} +
\frac{b_{21}\,r_0^2}{r^2}\right)\cos^2\theta\,,\\
\Sigma(r,\theta) &=& 1 +
\frac{A^2}{r^2}\cos^2\theta\,,\\
\label{eq:a4}
W(r,\theta) &=& \frac{1}{\Sigma(r,\theta)}\left[\frac{w_{00}\,r_0^2}{r^2} +
  \frac{w_{01}\,r_0^3}{r^3} + \left(\frac{w_{20}\,r_0^2}{r^2} +
  \frac{w_{21}\,r_0^3}{r^3}\right)\cos^2\theta\right]\,,\\
\label{eq:a5}
K^2(r,\theta) &=& 1 + \frac{AW(r,\theta)}{r} +
\frac{1}{\Sigma(r,\theta)}\left[\frac{k_{00}\,r_0^2}{r^2} +
  \left(\frac{k_{20}\,r_0^2}{r^2} +
  \frac{k_{21}\,r_0^3}{r^3}\right)\cos^2\theta\right]\,.
\end{eqnarray}
\end{widetext}

\begin{table*}
\begin{tabular}{|l|c|c|l|}
\hline
Parameter(s) & Constrained & Value & \hspace{5em} Description \\
\hline
$\epsilon_0$, $k_{00}$, $w_{00}$          & asymptotically & $\cdots$ & related to black-hole mass and angular momentum \\
\hline
$\epsilon_2$, $b_{20}$, $w_{20}$, $k_{20}$ & asymptotically & $0$ & for astrophysically realistic black holes \\
\hline
$a_{00}$, $b_{00}$ & asymptotically & $0$ & from current PPN estimates\\
\hline
$a_{20}$          & asymptotically & $\cdots$ & related to the quadrupole moment \\
\hline
$a_{01}$          & event horizon  & $\cdots$ & related to the deformation of $g_{tt}$\\
\hline
$a_{21}$, $k_{21}$ & event horizon  & $\cdots$ & related to the deformation of event horizon \\
\hline
$w_{01}$, $w_{21}$ & event horizon  & $\cdots$ & related to the rotational deformations of the metric \\
\hline
$b_{01}$, $b_{02}$ & event horizon  & $\cdots$ & related to the deformation of $g_{rr}$ \\
\hline
\end{tabular}
\caption{Summary of the properties of the various coefficients appearing
  in the lowest-order expression of the parametrized axisymmetric metric
  \eqref{eq:a1}--\eqref{eq:a5}. For each coefficient we report the regime
  where it is constrained, its value (when available), and the physical
  significance.}\label{Table3}
\end{table*}

In the expressions above, the values of the coefficients can be
constrained either by the asymptotic behavior of the metric (\ie
$\epsilon_0$, $k_{00}$, $w_{00}$, $\epsilon_2$, $b_{20}$, $w_{20}$,
$k_{20}$, and $a_{20}$), or by the conditions of the metric near the
black-hole event horizon (\ie $a_{01}$, $a_{21}$, $k_{21}$, $w_{01}$, $w_{21}$,
$b_{01}$, and $b_{21}$). More specifically, the three coefficients
$\epsilon_0$, $k_{00}$, and $w_{00}$ can be expressed in terms of radius
of the event horizon on the equatorial plane $r_0$, of the asymptotic
mass $M$, and of the rotation parameter $A$ as
\begin{equation*}
\epsilon_0=\frac{2M-r_0}{r_0}\,,\qquad
k_{00}=\frac{A^2}{r_0^2}\,, \qquad w_{00}=\frac{2MA}{r_0^2}\,.
\end{equation*}

In order to restore spherical symmetry at large distances at the first
PPN order, the coefficients $\epsilon_0$ and $b_{20}$ must vanish, while
the coefficient $k_{20}$ must vanish to maintain an asymptotic spherical
symmetry at the second PPN order, \ie in order to to have
\begin{equation}
g_{\phi\phi}=(r^2+a^2)\sin^2\theta+{\cal O}\left(\frac{1}{r}\right)
\,.
\end{equation}
Furthermore, if we exclude the rather exotic situation in
which the angular momentum of the black hole depends on the polar angle
$\theta$, \ie on the position of the observer relative to the equatorial
plane, then also the coefficient $w_{20}$ must vanish as well. As a
result, the following additional conditions can be imposed on the
coefficients $\epsilon_2$, $b_{20}$, $w_{20}$, and $k_{20}$ for
astrophysically realistic black-hole solutions \eqref{otherasympcoeff}
\begin{equation*}
\epsilon_2=0=b_{20}=w_{20}=k_{20}\,.
\end{equation*}
Similar considerations apply also to the coefficients $a_{00}$ and
$b_{00}$, which must vanish if one wants to match the first-order PPN
parameters of general relativity, \ie if $\beta=\gamma=1$ [\cf
Eqs. \eqref{a00} and \eqref{b00}]. Of course, these coefficients could be
taken to be nonzero if more exotic black-hole spacetimes are
investigated. Last but surely not least, the coefficient $a_{20}$ is
related to the black hole's quadrupole moment and is given
(\ref{quadrupole-moment})
\begin{equation*}
Q = -\frac{a_{20}\,r_0^3 + MA^2}{3}\,.
\end{equation*}

Next, we turn to the coefficients $a_{01}$, $a_{21}$, $b_{01}$, $b_{21}$,
$k_{21}$, $w_{01}$, and $w_{21}$, which describe the near-horizon
behavior of the metric. In particular, the coefficients $a_{21}$ and
$k_{21}$ describe deformations of the event horizon and, if the latter is
assumed to be spherical, must satisfy the condition
\begin{equation*}
a_{20}+a_{21}+k_{21}=0\,.
\end{equation*}
The coefficients $b_{01}$ and $b_{21}$, on the other hand, correspond to
deformations of the $g_{rr}$ metric function and are not expected to play
an important role in the dynamics of matter near the event
horizon\footnote{The properties of processes occurring on the equatorial
plane, such as those related to the position of ISCO or to the form of
the effective potential for particle motion (\ref{effective-potential}),
do not depend on the functions $B(r, \theta)$, and thus on the
coefficients $b_{01}$ and $b_{21}$.}.  Finally, the coefficients $w_{01}$
and $w_{21}$ are related to the rotational deformations of the metric,
while $a_{01}$ gives the PPN potential of the system. Table \ref{Table3}
offers a synthetic summary of the various properties of the coefficients,
of their values, and how they are constrained.

\subsection{Parametrized Kerr metric}
\label{sec:appendix_Kerr}

We first discuss the explicit form of the parametrized metric for a Kerr
black hole as obtained at first order in the radial direction and at
second order in the polar direction [\ie at $\mathcal{O}
(\cos^2\theta)$]. After setting $M$ and $A$ to be, respectively, the mass
and the specific angular momentum, \ie $A=a=J/M$, we obtain that the
event horizon is defined as
\begin{equation*}
r_0 = M+\sqrt{M^2-a^2}\,,
\end{equation*}
while the asymptotic coefficients have values
\begin{eqnarray*}
a_{00} &=& 0 = b_{00}\,,\\
a_{20} &=& \frac{2Ma^2}{r_0^3}\,.
\end{eqnarray*}
On the other hand, the strong-field coefficients are given by
\begin{eqnarray*}
a_{01} &=& 0 =w_{01}=w_{21}=b_{01}=b_{21}\,, \\
a_{21} &=& -\frac{a^4}{r_0^4}\,,\\
k_{21} &=& \frac{a^4}{r_0^4}-\frac{2Ma^2}{r_0^3}\,.
\end{eqnarray*}
Using this parametrization it is possible to reproduce the Kerr metric
in Boyer-Lindquist coordinates exactly.

\subsection{Parametrized dilaton black-hole metric}
\label{sec:appendix_dilbh}

Next, we turn to the explicit form of the parametrized metric for a
rotating dilaton black hole as obtained when truncating at first order in
the radial direction and at second order in the polar direction. After
setting
\begin{eqnarray*}
M &=& \mu+b\,, \\
A &=& a = J/M\,,
\end{eqnarray*}
the location of the event horizon is given by
\begin{eqnarray*}
r_0 &=& \sqrt{(\mu+b+\sqrt{\mu^2-a^2})^2-b^2} \,,\\
\end{eqnarray*}
while the first coefficients are given by
\begin{eqnarray*}
\epsilon_0 &=& \frac{2b+2\mu-r_0}{r_0}\,,\\
k_{00} &=& \frac{a^2}{r_0^2}\,,\\
w_{00} &=& \frac{2(b+\mu)a}{r_0^2}\,.
\end{eqnarray*}
The asymptotic coefficients are set to be
\begin{eqnarray*}
a_{00} &=& \frac{2b(b+\mu)}{r_0^2}\,,\\
b_{00} &=& 0\,,\\
a_{20} &=& \frac{2a^2(b+\mu)}{r_0^3}\,,
\end{eqnarray*}
while the strong-field ones are determined to be
\begin{eqnarray*}
a_{01} &=& \frac{2(\mu+b)\left[2b^2+r_0^2+(2r_0-3b)\sqrt{r_0^2+b^2}\right]}
{r_0^2\sqrt{r_0^2+b^2}} \\
&&-3\frac{r_0^2+a^2}{r_0^2}\,, \\
a_{21} &=& -\frac{a^4+2a^2(\mu+b)(b+r_0-\sqrt{r_0^2+b^2})}{r_0^4}\,,\\
k_{21} &=& \frac{a^4-2a^2(\mu+b)(\sqrt{r_0^2+b^2}-b)}{r_0^4}\,,\\
w_{01} &=& \frac{2a(\mu+b)(b+r_0-\sqrt{r_0^2+b^2})}{r_0^3}\,,\\
w_{21} &=& 0\,, \\
b_{01} &=& \frac{r_0}{\sqrt{r_0^2+b^2}}-1\,,\\
b_{21} &=& 0\,.
\end{eqnarray*}

\subsection{Parametrized EDGB black-hole metric}
\label{sec:appendix_EDGB}

Finally, we turn our attention to the parametrization of an EDGB black
hole as obtained when truncating at first order in the radial direction
and at second order in the polar direction. After setting $M$ to be the
mass $A=M\chi$, and taking into account that $\zeta =
16\pi\alpha^2/(\beta M^4)$, the position of the event horizon in the
equatorial plane is given by
\begin{equation*}
r_0 = 2M\left(1-\frac{\chi^2}{4}-\frac{49 \zeta }{80}+
\frac{128171 \chi^2\zeta}{588000}\right)+{\cal O}(\chi^4,\zeta^2)
\end{equation*}
and depends on $\zeta$. The asymptotic coefficients are given by
\begin{equation*}
a_{00} = b_{00} = 0\,,
\end{equation*}
and by
\begin{equation*}
a_{20} = \frac{\chi^2}{4}\left(1 + \frac{61429 \zeta}{14000}\right) + {\cal
  O}(\chi^4,\zeta^2)\,,
\end{equation*}
thus implying violation of the Kerr expression for the quadrupole moment
[\cf Eq. \eqref{EDGBquadruplole}]. Finally, the strong-field coefficients
are found to be
\begin{eqnarray*}
a_{01} &=& -\frac{17 \zeta }{60}\left(1-\frac{324899
  \chi^2}{166600}\right)+{\cal O}(\chi^4,\zeta^2)\,,\\
a_{21} &=& \frac{447731}{392000}\chi^2\zeta+{\cal O}(\chi^4,\zeta^2)\,,\\
k_{21} &=& -\frac{\chi^2}{4}\left(1+\frac{438867
  \zeta}{49000}\right)+{\cal O}(\chi^4,\zeta^2)\,,\\
w_{01} &=& -\frac{63\chi\zeta}{160}+{\cal O}(\chi^3,\zeta^2)\,,\\
w_{21} &=& {\cal O}(\chi^3,\zeta^2)\,,\\
b_{01} &=& -\frac{361 \zeta }{240}\left(1-\frac{51659
  \chi^2}{176890}\right)+{\cal O}(\chi^4,\zeta^2)\,,\\
b_{21} &=& \frac{175629}{196000}\chi^2\zeta+{\cal O}(\chi^4,\zeta^2)\,.
\end{eqnarray*}

\end{document}